\documentclass[journal=jacsat,manuscript=article]{achemso}
\SectionNumbersOn

\usepackage[utf8]{inputenc}

\usepackage{amsmath}
\usepackage{color}
\usepackage{graphicx}
\usepackage{comment}
\usepackage{caption}
\usepackage{subcaption}
\usepackage[normalem]{ulem}
\usepackage{algorithm}
\usepackage{algpseudocode}

\usepackage{rotating}
\usepackage{amsmath,amssymb}
\usepackage{graphicx}
\usepackage{booktabs}
\usepackage{xcolor}

\def\boldr{{\mathbf{r}}}
\def\boldrp{{\mathbf{r}'}}
\def\br{{\mathbf{r}}}
\def\brp{{\mathbf{r}^{\prime}}}
\author{Johannes T\"{o}lle}
\altaffiliation{Both authors contributed equally.}
\affiliation{Division of Chemistry and Chemical Engineering, \\
California Institute of Technology, Pasadena, California 91125, USA}
\email{jtolle@caltech.edu}

\author{Niklas Niemeyer}
\altaffiliation{Both authors contributed equally.}

\author{Johannes Neugebauer}
\affiliation{University of M\"unster, Organisch-Chemisches Institut and \\
Center for Multiscale Theory and Computation,\\
Corrensstra{\ss}e 36, 48149 M\"unster, Germany}
\email{j.neugebauer@uni-muenster.de}

\title{Accelerating Analytic-Continuation GW Calculations with a Laplace Transformation and Natural Auxiliary Functions}

\begin{document}

\begin{abstract}
We present a simple and accurate GW implementation based on a combination of a Laplace transformation (LT) and other acceleration techniques used in post-SCF quantum chemistry, namely, natural auxiliary functions and the frozen-core approximation.
The LT-GW approach combines three major benefits: (a) a small prefactor for the computational scaling, (b) easy integration into existing molecular GW implementations, and (c) significant performance improvements for a wide range of possible applications. 
Illustrating these advantages for systems consisting of up to 352 atoms and 7412 basis functions, we further demonstrate the benefits of this approach combined with an efficient implementation of the Bethe--Salpeter equation.
    
\end{abstract}
\clearpage

\section{Introduction}
After its introduction in 1965 \cite{hedin1965new}, the GW (G: time ordered one-body Green’s function, W: screened
Coulomb interaction)  method has now become the standard approach for the accurate \textit{ab-initio} determination of ionization potentials (IPs), electron affinities (EAs) (or more generally quasi-particle energies), and in combination with the Bethe--Salpeter equation (BSE), for excitation energies in condensed matter physics \cite{onida1995ab,rohlfing1998excitonic,albrecht1998ab,rohlfing1998electron,benedict1998optical,rohlfing2000electron,baumeier2012excited}.
The adoption within the realm of quantum chemistry has been established in recent years~\cite{ren2012resolution,van2013gw,jacquemin2015benchmarking,bruneval2016molgw,wilhelm2016gw,krause2017implementation,golze2018core,2020turbomole,foerster2020low,liu2020all,zhang2023many} with the availability of implementations in a wide range of molecular quantum chemistry codes, see e.g., Refs.~\citenum{bruneval2016molgw,fiestprog,2020turbomole,sun2020recent,zhu2021all,unsleber2018,niemeyer2022subsystem,serenity152,toelle2021,forster2021low,ren2012resolution,tirimbo2020excited,apra2020nwchem,mejia2021scalable}.
The success of the GW method is owed to the fact that it offers good accuracy while being computationally feasible for a wide range of systems, c.f.~Ref.~\citenum{bruneval2021gw}.
However, the GW method generally relies on error cancellation, and G$_0$W$_0$, in particular, depends on the starting point chosen, the approach used for determining the dielectric function, and the self-consistency scheme chosen for the GW calculation.
An excellent overview of the different aspects related to the GW approximation can be found in Ref.~\citenum{golze2019gw}.

Especially the computational cost for determining the screened Coulomb interaction and therefore the G$_0$W$_0$ self-energy $\Sigma_0$ varies significantly for different practical realizations of the GW method in molecular orbital bases.
The ``fully-analytic'' approach \cite{sett2012,bruneval2012ionization}, for example, scales as $\mathcal{O}(N^6)$.
The scaling can be reduced significantly by numerical integration of the self-energy $\Sigma_0$,
\begin{align}
        \Sigma_0(\boldr,\boldrp,\omega) = \frac{\mathrm{i}}{2\pi}\int d\omega' e^{\mathrm i\omega'\eta} G_0(\boldr,\boldrp,\omega+\omega') W_0(\boldr,\boldrp,\omega'),
        \label{eq:SelfEnergy}
\end{align}
where the non-interacting one-particle Green's function is denoted as $G_0$ and the screened Coulomb interaction as $W_0$. 

To avoid divergences along the real frequency axis \cite{golze2018core}, the integration in Eq.~(\ref{eq:SelfEnergy}) is commonly performed along the imaginary frequency axis in combination with analytic continuation (AC) to the real frequency axis leading to a formal scaling of 
$\mathcal{O}(N^4)$ \cite{wilhelm2016gw,wilhelm2018toward,zhu2021all}.
Alternatively, one can employ the so-called contour-deformation approach (CD) \cite{godby1988self,holzer2019ionized,golze2018core,zhu2021all} by dividing the integration in Eq.~(\ref{eq:SelfEnergy}) into an integration along the imaginary frequency axis and the real-frequency axis.
The scaling, however, is $\mathcal{O}(N^{4-5})$ and
depends on the quasi-particles to be determined (see Ref.~\citenum{golze2018core}).

$\Sigma_0$ can also be determined within the space-time formulation of the GW method~\cite{rojas1995space,liu2016cubic,wilhelm2018toward,wilhelm2021low,forster2021gw100,duchemin2021cubic}.
In this approach, the construction of $W_0$ is performed in imaginary-time rather than frequency space in combination with additional techniques, among others, real-space grid representation of the Green's function \cite{liu2016cubic,kaltak2014low}, pair atomic density fitting \cite{foerster2020low}, or separable density-fitting \cite{duchemin2019separable,duchemin2021cubic} to reduce the overall scaling to $\mathcal{O}(N^3)$ which allows for its application to systems containing almost 1000 atoms, e.g.~Ref.~\citenum{wilhelm2021low}.
Note that this ansatz is equivalent to Laplace-transform (LT)
techniques used in molecular quantum chemistry~\cite{almlof1991,haser1992,haser1993}.
Drawbacks of these methods are, however, related to increasing memory requirements and larger prefactors due to the real-space representation \cite{liu2016cubic}, the careful error evaluation necessary concerning the various numerical procedures and chosen cut-offs \cite{foerster2020low,wilhelm2021low}, or the necessity to construct specialized real-space grids \cite{duchemin2021cubic}.
These aspects also lead to more challenging numerical implementations of these methods, potentially limiting their widespread application. Note, however, that in the limit of very large systems, these approaches can be more beneficial compared to the methodology presented here.

This work demonstrates an alternative efficient evaluation of the GW self-energy by combining different ideas for reducing the computational cost based on the AC-GW formulation. In particular, we make use of a Laplace transformation for the evaluation of $W_0$, a truncation of the auxiliary basis using natural auxiliary functions (NAF) \cite{kallay2014,mester2017} and the frozen-core (FC) approximation.
We refer to this approach as LT-GW in the following which is based on three guiding principles: 
(a) a small prefactor should be preserved, (b) adaptation of existing AC-GW implementations should require minimal effort, and (c) significant performance improvements should result for a wide range of system sizes with controllable error.

\section{Theory}
In the following, a concise overview of the modified GW implementation based on the Laplace-transform (LT) technique is given. More detailed information regarding GW implementations based on imaginary frequency integration can be found in Refs.~\citenum{golze2018core,holzer2019ionized,zhu2021all}.

A diagonal element $nm$ for the correlation part of the screened-Coulomb interaction $W^c_{nm}$ in a molecular orbital basis for an imaginary frequency $\mathrm{i}\omega$ is calculated as 
\begin{equation} \label{eq:ScreenedCoulombOrbital}
        W^c_{nm}(\mathrm{i}\omega') = \sum_{PQ} R^P_{nm}\left\{\left[\mathbf{1} - \mathbf{\Pi}(\mathrm{i}\omega')\right]_{PQ}^{-1} - \delta_{PQ} \right\}R^Q_{nm},
\end{equation}
where molecular spin-orbital ($\phi$) and auxiliary basis function ($\chi$) indices are given in lowercase and uppercase letters, respectively. 
Furthermore, $i,j,\dots$ refer to occupied, $a,b,\dots$ to virtual, and $n,m,\dots$ to arbitrary orbitals with eigenvalues $\epsilon$. 
$\Pi_{PQ}(\mathrm i\omega')$ is evaluated as 
\begin{align}
        \Pi_{PQ}(\mathrm i\omega') = - 2 \sum_{ia}R^P_{ia} \frac{\left(\epsilon_a - \epsilon_i\right)}{\omega'^2 + \left(\epsilon_a - \epsilon_i\right)^2} R^Q_{ia},
        \label{eq:Pi}
\end{align}
and the transformed three-center integrals $R^P_{nm}$ are defined as 
\begin{align}
    R^Q_{nm} = \sum_P (nm|P) [\mathbf{V}^{-1/2}]_{PQ},
\end{align}
with 
\begin{align}
    (nm|P) = \int d\br \int d\brp \frac{\phi_n(\br) \phi_m(\br) \chi_P(\brp)}{|\br - \brp|},
\end{align}
and 
\begin{align}
        V_{PQ} = \int d\br \int d\brp \frac{\chi_P(\br) \chi_Q(\brp)}{|\br-\brp|}.
\end{align}
In AC-GW, the construction of $\Pi_{PQ}(\mathrm i\omega')$ is the most time-consuming step, formally scaling as $\mathcal{O}(N_\mathrm{o}N_\mathrm{v}N_\mathrm{aux}^{2})$ for each imaginary frequency ($N_\mathrm{o}$ being the number of occupied orbitals, $N_\mathrm{v}$ the number of virtual orbitals, and $N_\mathrm{aux}$ the number of auxiliary functions). 
Finally, the correlation (dynamical) part of the G$_0$W$_0$ self-energy $\Sigma^c$ is obtained ($\epsilon_F$ denotes the Fermi-level)
\begin{align}
        &\Sigma_n^c(\mathrm{i}\omega)=  -\frac{1}{\pi} \sum_m \int_{0}^\infty d \omega' \frac{\mathrm{i}\omega + \epsilon_F - \epsilon_m}{\left(\mathrm{i}\omega + \epsilon_F - \epsilon_m \right)^2 + \omega'^2} W_{nm}(\mathrm{i}\omega'),
        \label{eq:CorrelationSelfEnergy}
\end{align}
which is integrated numerically using a modified Gauss--Legendre (GL) quadrature, see Refs.~\citenum{ren2012resolution,zhu2021all}.
We note that different integration choices for Eq.~(\ref{eq:CorrelationSelfEnergy}) are possible. 
In this work, however, we restrict ourselves to a modified GL quadrature since it presumably is the most common numerical integration scheme used in the analytic-continuation GW context.
Quasi-particle energies are then determined by AC of $\Sigma^c$ to the real frequency axis.
For the AC to the real frequency axis, we use a $N$-point Pad\'{e} approximation as described in the appendix of Ref.~\citenum{vidberg1977solving}. 

In this work, we make use of the LT for evaluating $\Pi_{PQ}(\mathrm i\omega')$.
In a first step, the denominator in Eq.~(\ref{eq:Pi}) is rewritten as 
\begin{align}
    \frac{1}{\omega'^2 + (\epsilon_a - \epsilon_i)^2} &= \int^\infty_0 d\tau \exp(-\left(\omega'^2 + (\epsilon_a - \epsilon_i)^2\right)\tau) \nonumber \\
    & = \int^\infty_0 d\tau \exp\left(-\omega'^2\tau\right) \exp(-( \epsilon_a - \epsilon_i)^2 \tau).
    \label{eq:LT}
\end{align}
holding for $\left(\omega'^2 + (\epsilon_a - \epsilon_i)^2\right) > 0$ which is guaranteed to be true.
Replacing the denominator with the integral in Eq.~(\ref{eq:LT}) allows to apply a numerical integration of the form
\begin{align}
    \frac{1}{\omega'^2 + (\epsilon_a - \epsilon_i)^2}&\approx - \sum_m^{N_\mathrm{LT}}  w_m \exp\left(-\left(\omega'^2 + (\epsilon_a - \epsilon_i)^2\right) x_m\right) \nonumber \\
    &= - \sum_m^{N_\mathrm{LT}}  w_m \exp\left(-\omega'^2 x_m\right) \exp\left(-(\epsilon_a - \epsilon_i)^2 x_m\right),
\end{align}
where the $N_\mathrm{LT}$ quadrature points and their corresponding weights are denoted as $x_m$ and $w_m$, respectively.
Factorizing the exponential functions with frequencies and orbital-energy differences
as their arguments through the LT allows evaluating their contributions to $\Pi_{PQ}(\mathrm i\omega')$ separately as
\begin{equation}
    \Pi_{PQ}(\mathrm i\omega') \approx -2 \sum_m \underbrace{\sum_{ia}R^P_{ia} w_m \left(\epsilon_a - \epsilon_i\right) e^{-(\epsilon_a - \epsilon_i)^2 x_m}   R^Q_{ia}}_{M^m_{PQ}} e^{-\omega'^2 x_m}.
\end{equation}
In practice, $M^m_{PQ}$ is calculated for each quadrature point, which requires $N_\mathrm{LT} N_\mathrm oN_\mathrm vN_\mathrm {aux}^{2}$ operations, followed by the outer loop over imaginary frequencies [see Eq.~(\ref{eq:ScreenedCoulombOrbital})] counting $N_\mathrm{LT} N_\mathrm {aux}^{2} N_{\mathrm{i}\omega}$ operations.
In contrast, the evaluation of Eq.~(\ref{eq:Pi}) for the determination of quasi-particle energies requires $N_{\mathrm{i}\omega} N_\mathrm oN_\mathrm vN_\mathrm{aux}^{2}$ operations.
It becomes clear that the formal scaling remains unchanged with $\mathcal{O}(N^4)$ since neither
$N_{\mathrm{i}\omega}$ nor $N_\mathrm{LT}$ depends on the system size represented by $N$. A constant speed-up can, however, be expected using the LT technique as long as $N_\mathrm{LT} < N_\mathrm{\mathrm{i}\omega}$ which is proportional to the ratio $N_{\mathrm{i}\omega}/N_\mathrm{LT}$.

The natural auxiliary function (NAF) approximation~\cite{kallay2014} reduces the size of the three-index integral tensor that commonly appears in post-SCF methodology making use of the resolution of the identity approximation. Its basis is given by a symmetric, positive definite matrix $K$ that reads
\begin{equation}\label{nafeq}
  K_{PQ} = \sum_{nm} R^P_{nm}R^Q_{nm}.
\end{equation}
A rank reduction of the three-index integral list is achieved by first diagonalizing $K$ to yield the NAFs labeled by $\tilde P$, 
\begin{equation}
 \sum_Q K_{PQ}  V_{Q,\tilde P} = V_{P \tilde P} \epsilon_{\tilde P} ,
\end{equation}
followed by setting up a transformation matrix $U_{P\tilde P}$ that only includes NAFs with corresponding eigenvalues above a certain threshold $\varepsilon_\text{NAF}$ (assembled from the columns of $V_{P \tilde P}$). Finally, the three-center integral
tensor is transformed to the NAF space following
\begin{equation} \label{eq:naftrafo}
  R^{\tilde{P}}_{nm} = \sum_{P} R^P_{nm} U_{P\tilde P}.
\end{equation}
In the limit of $U$ including \emph{all} eigenvectors of $K$, Eq.~(\ref{eq:naftrafo}) represents an orthogonal transformation. 
Our implementation omits the virtual--virtual part of the sum in Eq.~(\ref{nafeq}) due to its unfavorable scaling with the system size. 
Closed-shell molecules are handled by including a factor of two in Eq.~(\ref{nafeq}) to account for the single set of spatial orbitals.
Determining the NAFs formally scales as $\mathcal{O}(N_\mathrm{o} N_\mathrm{v} N^2_\mathrm{aux})$. The theoretical speed-up of the NAF approximation in AC-GW calculations becomes apparent when inspecting Eqs.~(\ref{eq:Pi}) and~(\ref{eq:LT}).  The time-determining step includes an inner product of the three-index integral tensor contracting the occupied--virtual composite index $ia$.
As a result, the expected speed-up scales quadratically with the quotient of the number of original auxiliary basis functions $N_\mathrm{aux}$ and the number of NAFs $N_\mathrm{NAF}$, that is, $(N_\mathrm{aux}/N_\mathrm{NAF})^2$.

\section{Computational Details}
\label{sec:compdetails}
All calculations presented in this article were performed with a slightly modified version of the \textsc{Serenity}
program (1.5.2)~\cite{unsleber2018,niemeyer2022subsystem,serenity152}. All self-consistent field (SCF) procedures were stopped as soon as two of the following
convergence criteria have been met: total energy threshold of $5\cdot 10^{-8}$ E$_\mathrm{h}$, root-mean-square deviation of the density matrix threshold of $5\cdot 10^{-8}$ a.u., as well as a threshold of $5\cdot 10^{-7}$ a.u. for the commutator of the Fock and density matrix.
DFT calculations employ default grids as implemented in \textsc{Serenity}.
All calculations employ def2-TZVP basis sets \cite{weigend2005balanced} and the GW and BSE calculations additionally use the corresponding RIFIT (RI-C) basis set~\cite{weigend1998ri}. Unless explicitly stated otherwise, the resolution of the identity approximation is applied to the Coulomb part of the
Fock matrix in SCF calculations with the universal def2/JFIT basis set~\cite{weigend2006accurate}.
GW calculations were performed within the analytic-continuation (AC) approach, include the energetically lowest and highest five virtual and occupied SCF orbitals, unless stated otherwise, and employ 128 integration points along the imaginary frequency axis that were obtained from a modified Gauss--Legendre (GL) quadrature.
The Pad\'{e} approximation is performed based on 70 imaginary frequencies obtained from a modified Gauss--Legendre grid\cite{ren2012resolution,zhu2021all}.
Bethe--Salpeter equation (BSE) calculations were carried out within the Tamm--Dancoff and static (to the dielectric function) approximations.
Eigenvectors in the iterative solution of the BSE were converged to maximum residual norms of $10^{-5}$ and GW quasi-particle energies are used for the construction of the static dielectric function.
Orbitals not included in the quasi-particle calculation are shifted 
based on the difference between the lowest and highest quasi-particle energy
and the respective KS orbital eigenvalue \cite{holzer2019ionized,toelle2021}.
We used six quasi-particle iterations for the water cluster, discussed in Fig.~1 and 2, and otherwise, quasi-particle iterations are performed until the change in the HOMO/LUMO gap is below $10^{-6}~\mathrm{E_h}$.
For calculations related to the GW100 benchmark, the imaginary broadening factor $\eta$ is set to $0.0$ for comparison purposes, while it is set to $\eta =  0.001$~a.u. for the remaining calculations.
Quadrature points and weights for the Laplace transformation were obtained with the \texttt{laplace-minimax} library~\cite{takatsuka2008,helmich2016}.
For this, the lower bound of the denominator was chosen to be the squared difference of the highest occupied molecular orbital (HOMO) eigenvalue and the lowest unoccupied molecular orbital (LUMO) eigenvalue summed with the square of the smallest frequency.
The upper bound was chosen to be the square of the largest energy difference of the energetically lowest occupied and highest unoccupied molecular orbital eigenvalues.
If this upper bound is below $10^{4}$ a.u., we use $10^{4}$ a.u.~instead to ensure
an accurate numerical integration for a wide energy window.
As shown in this article, these bounds together with the chosen threshold $\varepsilon_\mathrm{LT} = 10^{-7}$ for the square root of the error function of the LT procedure lead to negligible errors in the quasi-particle energies ($<0.1$ meV). We note that this threshold is far tighter than commonly used thresholds in, e.g., coupled-cluster applications, where $10^{-3}$--$10^{-4}$ is a standard choice~\cite{winter2011}.
In summary, all thresholds for numerical integration (LT-/modified Gauss--Legendre-grid) as well as the number of frequencies for the Pad\'e approximation in the AC procedure are chosen very conservatively in order to keep the introduced error as small as possible. An analysis of the influence of the size of the LT and Gauss--Legendre grid on the accuracy and computational timings in the determination of QP energies are discussed in Sec.~\ref{sec:waterclusters} for a system consisting of 100 water molecules.
Within the frozen-core (FC) approximation, a tabulated number of the energetically lowest-lying SCF orbitals for each atom are frozen in all post-SCF treatments. These numbers were chosen following the defaults in the ORCA program (as listed in Ref.~\citenum{orcaprog}).

\section{Results}
\subsection{Quasi-particle energies using LT-G$_0$W$_0$}
\subsubsection{GW100}
\begin{figure}[!b]
    \centering
    \includegraphics[width=\linewidth]{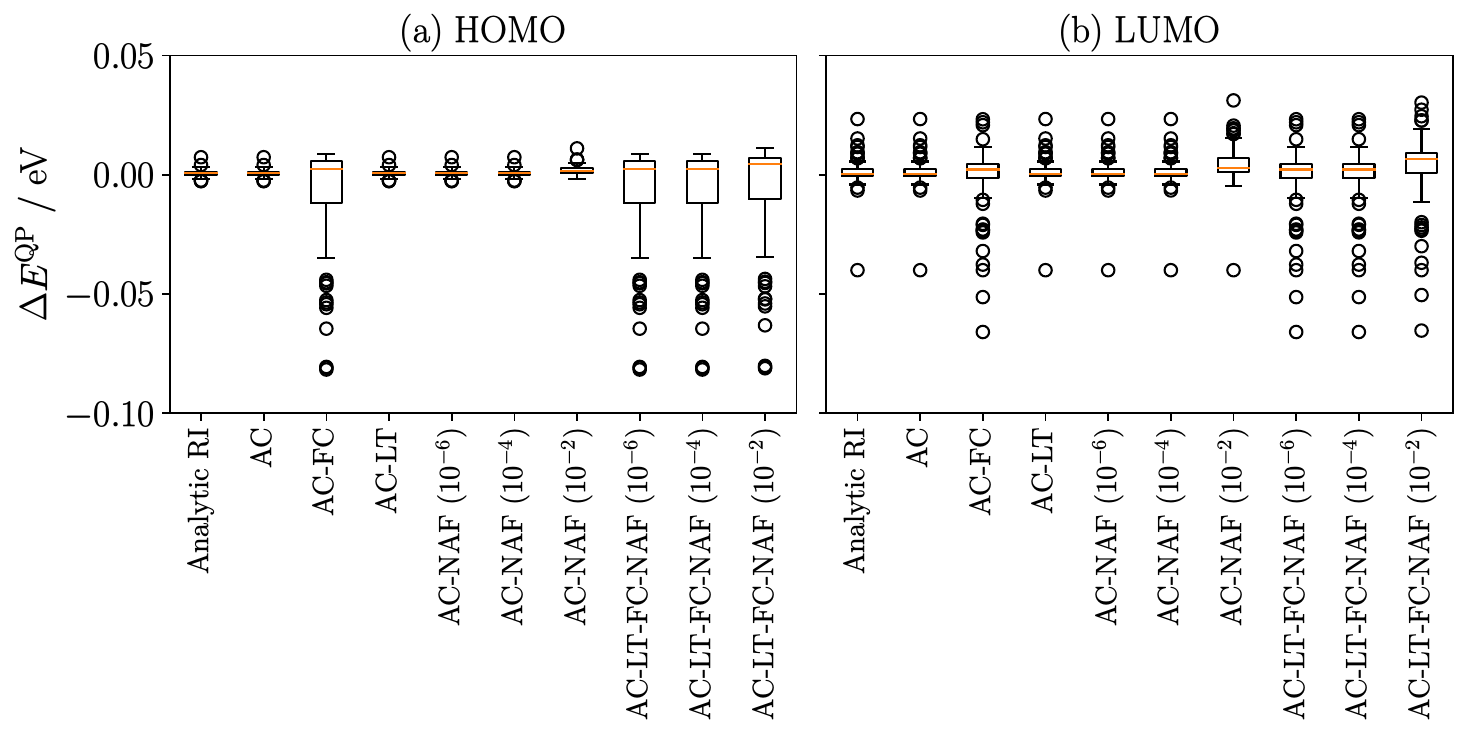}
    \caption{Deviations in the G$_0$W$_0$ (a) HOMO and (b) LUMO quasi-particle energies for the molecular systems from the GW100 benchmark set for various approximations in the evaluation of the G$_0$W$_0$ self-energy relative to the ``fully-analytic'' approach without employing the RI approximation. Analytic-RI: ``Fully-analytic'' approach using the RI approximation, AC: AC-G$_0$W$_0$, LT: AC-G$_0$W$_0$ in combination with AC-LT ($\varepsilon_\text{LT}=10^{-7}$), AC-FC: AC-G$_0$W$_0$ in combination with FC, AC-NAF: AC-G$_0$W$_0$ in combination with the NAF approximation ($\varepsilon_\text{NAF} = 10^{\{-6,-4,-2\}}$), AC-FC-LT-NAF: Combining AC-G$_0$W$_0$ with FC/LT/NAF ($\varepsilon_\text{LT}=10^{-7}$, $\varepsilon_\text{NAF} = 10^{\{-6,-4,-2\}}$) [def2-TZVP, starting from Hartree--Fock orbitals].
    The box plots were created with the \texttt{matplotlib.pyplot.boxplot} function of the \texttt{matplotlib} library using default settings. For each data set, the orange line marks the median, the top and bottom of the box mark the 25$^\mathrm{th}$ and 75$^\mathrm{th}$ percentiles, respectively, and their difference, i.e. the box height is the interquartile range (IQR). The lower and upper ends of the whiskers mark the lowest and highest value before the 25$^\mathrm{th}$ percentile minus and 75$^\mathrm{th}$ percentile plus, respectively, one and a half times the IQR. Circles mark values outside of this region, generally considered outliers. For further explanation of the different elements of the box plot, see Ref.~{\protect\citenum{matplotlib}}. The HOMO and LUMO quasi-particle energies themselves are found in Tabs.~S1 and S2.}
    \label{fig:DeviationBoxplot}
\end{figure}

In the following, we will demonstrate the robustness, scalability, and speed-up of combining AC-G$_0$W$_0$ with the LT, NAF, and FC techniques.
First, its accuracy is determined for the GW100 benchmark set~\cite{van2015gw}. Reference orbitals were obtained using the Hartree--Fock approximation throughout. Effective core potentials are used for the heavy elements rubidium, silver, xenon, and iodine.
All results are compared to reference quasi-particle (QP) energies based on the ``fully-analytic'' evaluation of the G$_0$W$_0$ self-energy without employing the RI approximation (also for the mean-field calculation) \cite{van2013gw}.
The resulting deviations for the HOMO/LUMO quasi-particle energies are displayed in Fig.~\ref{fig:DeviationBoxplot}.
Statistical measures [mean absolute error (MAE), maximum absolute error (MXAE) and standard deviation (SD)] are given in Tab.~\ref{tab:GW100Statistics}.
The quasi-particle energies for all methods are explicitly shown in Tab.~S1 (HOMO) and Tab.~S2 (LUMO) in the Supporting Information (SI).

From Fig.~\ref{fig:DeviationBoxplot} it becomes clear that especially for the HOMO quasi-particle energies evaluated with analytic-continuation (AC) in combination with the frozen-core (FC) approximation, larger deviations up to $-0.08$ eV for the HOMO quasi-particle energies of, for example, krypton, bromine and carbon tetrabromide are observed.
These systems also show comparatively large errors for the LUMO quasi-particle energies of $-0.07$ eV and $-0.02$ eV, respectively.
\begin{table}[!t]
    \centering
    \caption{Mean absolute error (MAE), maximum absolute error (MXAE) and standard deviation (SD) in the G$_0$W$_0$ (a) HOMO and (b) LUMO quasi-particle energies for the molecular systems from the GW100 benchmark set for various approximations in the evaluation of the G$_0$W$_0$ self-energy relative to the ``fully-analytic'' approach without employing the RI approximation in meV. Analytic-RI: ``Fully-analytic'' approach using the RI approximation, AC: AC-G$_0$W$_0$, LT: AC-G$_0$W$_0$ in combination with AC-LT ($\varepsilon_\text{LT}=10^{-7}$), AC-FC: AC-G$_0$W$_0$ in combination with FC, AC-NAF: AC-G$_0$W$_0$ in combination with the NAF approximation ($\varepsilon_\text{NAF} = 10^{\{-6,-4,-2\}}$), AC-FC-LT-NAF: Combining AC-G$_0$W$_0$ with FC/LT/NAF ($\varepsilon_\text{LT}=10^{-7}$, $\varepsilon_\text{NAF} = 10^{\{-6,-4,-2\}}$) [def2-TZVP, starting from Hartree--Fock orbitals].}
    \label{tab:GW100Statistics}
    \begin{tabular}{lrrr | rrr}
    \toprule
     & \multicolumn{3}{c}{(a) HOMO} & \multicolumn{3}{c}{(b) LUMO} \\
     \hline
     & MAE & MXAE & SD & MAE & MXAE & SD\\
     \hline
    Analytic RI & 1.1 & 7.3 & 1.4 & 2.9 & 40.0 & 5.8 \\ 
    AC & 1.1 & 7.3 & 1.4 & 2.9 & 40.0 & 5.8 \\ 
    AC-FC & 13.3 & 81.7 & 21.7 & 7.9 & 65.9 & 13.4 \\ 
    AC-LT & 1.1 & 7.3 & 1.4 & 2.9 & 40.0 & 6.0 \\ 
    AC-NAF ($10^{-6}$) & 1.1 & 7.3 & 1.4 & 2.9 & 40.0 & 5.8 \\ 
    AC-NAF ($10^{-4}$) & 1.1 & 7.3 & 1.4 & 2.9 & 40.0 & 5.8 \\ 
    AC-NAF ($10^{-2}$) & 2.1 & 11.1 & 1.9 & 5.6 & 40.0 & 7.4 \\ 
    AC-LT-FC-NAF ($10^{-6}$) & 13.3 & 81.7 & 21.7 & 7.9 & 65.9 & 13.4 \\ 
    AC-LT-FC-NAF ($10^{-4}$) & 13.3 & 81.7 & 21.7 & 7.9 & 65.9 & 13.4 \\ 
    AC-LT-FC-NAF ($10^{-2}$) & 14.0 & 81.1 & 22.0 & 10.8 & 65.3 & 14.8 \\
    \bottomrule
    \end{tabular}
\end{table}
These errors transfer to the quasi-particle energies evaluated within the AC/FC/NAF approximation.
Because these errors originate from the FC approximation, we investigated the deviations in the quasi-particle energies of these systems by reducing the number of frozen electrons from 18 $e^{-}$ to 10 $e^{-}$.
The resulting quasi-particle deviations are shown in Tab.~\ref{tab:FC_benchmark}.
It can be seen that the errors with the modified FC are below 4 meV, which highlights that the FC can be systematically adjusted to reduce the resulting error in the quasi-particle energies.
All systems studied beyond the GW100 benchmark set contain only first- and second-row elements (with WW-6 being an exception, which is, however, separately benchmarked against a regular AC-G$_0$W$_0$ calculation).
For these systems, the FC approximation leads to only a small error.
We, therefore, used the default number of frozen electrons in the remaining calculations.
For the LUMO quasi-particle energies, we find that the deviation of the hydrogen system is much larger than for the rest of the systems already for the ``Analytic RI'' calculation (which transfers directly to all AC calculations, see the outlier in Fig.~\ref{fig:DeviationBoxplot}(b)). We can, therefore, conclude that these deviations are a result of the RI approximation for this system and that the LT, NAF, or FC approximations introduce only small and controllable errors.
These deviations are much smaller than the intrinsic error of the G$_0$W$_0$ method itself, c.f.~Ref.~\citenum{bruneval2021gw}, justifying their application.
Especially the loose NAF threshold of $10^{-2}$ leads to almost negligible error. As a result, all further calculations shown in this article will be confined to this threshold.
\begin{table}[!t]
    \caption{Deviations in the HOMO/LUMO G$_0$W$_0$ quasi-particle energies for AC-G$_0$W$_0$ in combination with the FC approximation (relative to the ``fully-analytic'' approach without employing the RI approximation), either freezing 18$e^{-}$ or 10$e^{-}$ of bromine, krypton, and carbon tetrabromide. The number of frozen carbon electrons remains unchanged [def2-TZVP, starting from Hartree--Fock orbitals].}
    \label{tab:FC_benchmark}
    \centering
    \begin{tabular}{l | rr | rr}
    \toprule
    Molecule & \multicolumn{2}{c |}{FC(18 $e^{-}$)} & \multicolumn{2}{c}{FC(10 $e^{-}$)} \\
    & HOMO & LUMO & HOMO & LUMO \\
    \midrule
    bromine & $-$0.0806 & $-$0.0659 &  $-$0.0037 & 0.0034 \\
    krypton & $-0.0809$ & $-0.0230$ & $-$0.0020	& 0.0000	 \\
    carbon tetrabromide & $-0.0817$ & $-0.0377$ & $-0.0037$ & $0.0010$ \\
    \bottomrule 
    \end{tabular}
\end{table}

\subsubsection{Water clusters}
\label{sec:waterclusters}
\begin{figure}[!t]
    \centering
    \includegraphics[width=0.85\linewidth]{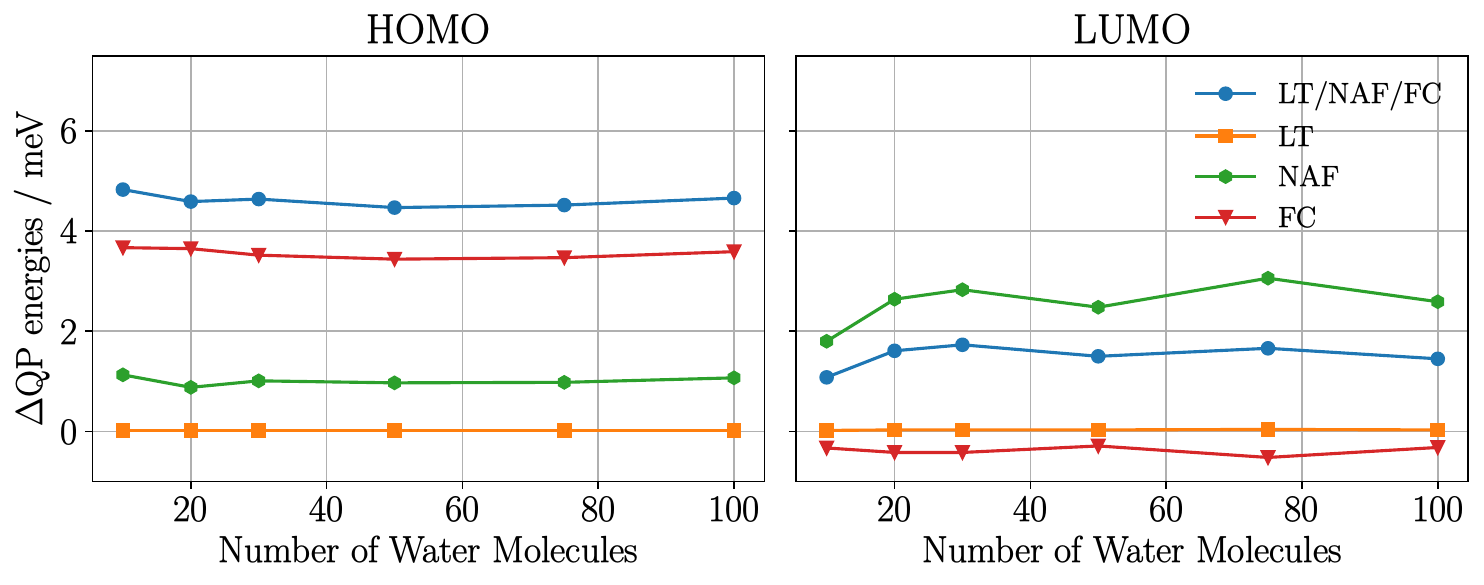}
    \caption{Signed error in the HOMO and LUMO G$_0$W$_0$ QP energies as a function of the number of molecules included in the water cluster shown in Fig.~{\protect\ref{fig:doublelogplot}} [HF/def2-TZVP].}
    \label{fig:qperrorwater}
\end{figure}

\begin{figure}[!t]
    \centering
    \includegraphics[width=0.95\linewidth]{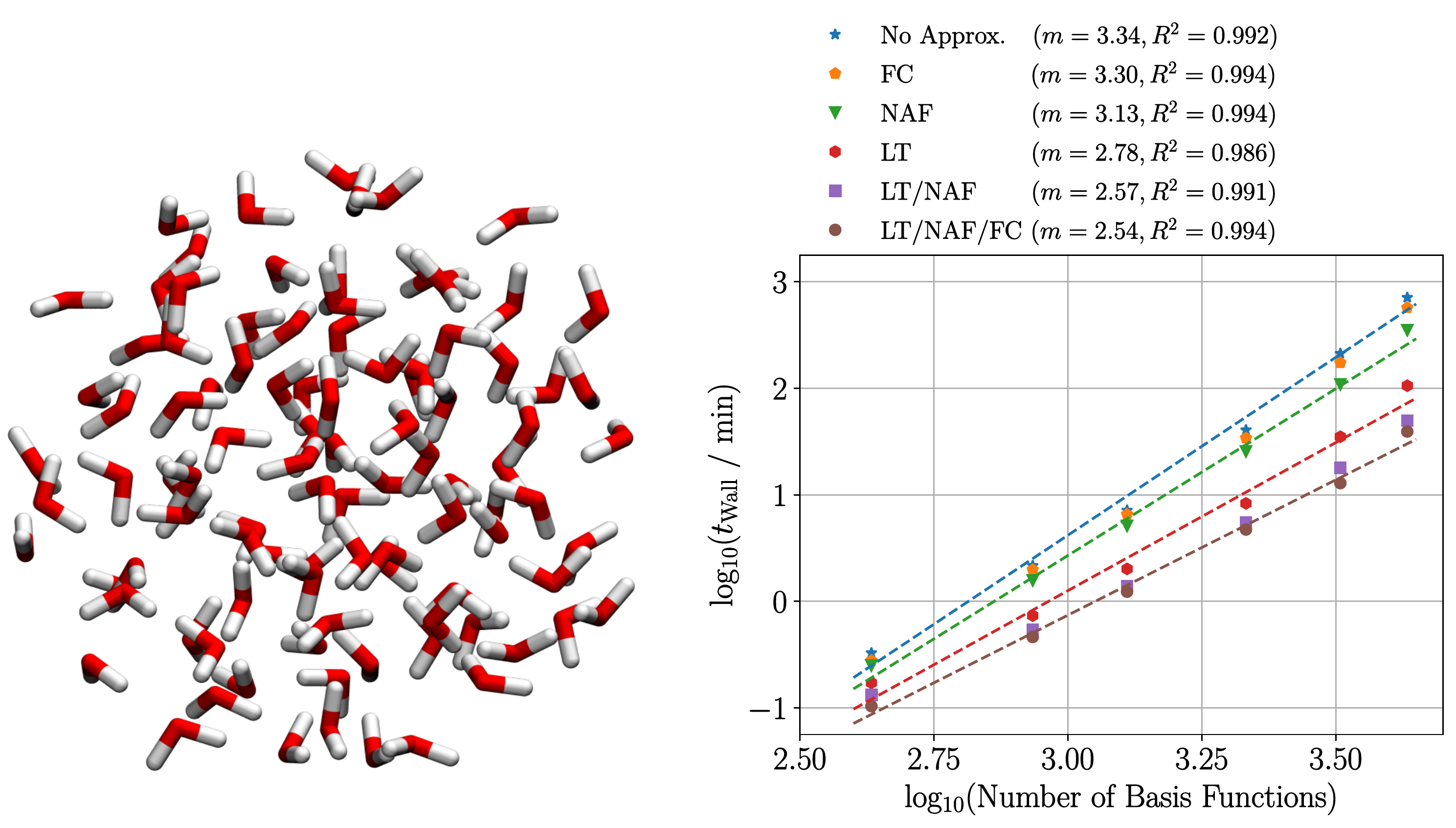}
    \caption{Left: Water cluster containing 100 water molecules. Right: Wall-clock timings as a function of the number of SCF basis functions (double-logarithmic plot) as well as slopes $m$ and coefficients of determination $R^2$ of linear fit functions.}
    \label{fig:doublelogplot}
\end{figure}

Next, we performed G$_0$W$_0$ calculations on water clusters (see Fig.~\ref{fig:doublelogplot}) of increasing size containing ten to 100 water molecules (corresponding to 430 to 4300 SCF basis functions in a def2-TZVP basis, respectively) and investigate QP energies and computational timings. The geometries were obtained by first generating a cubic $20\times 20\times 20$~\AA$^3$~water cluster containing 233 water molecules with VMD~\cite{humpf1996}, optimizing it with GFN2-xTB~(6.4.1)~\cite{bannwarth2019gfn2} and then including the respective number of molecules closest to the center of mass of the whole cluster.
In Fig.~\ref{fig:qperrorwater}, we display the signed error in QP energies as a function of the number of molecules included in the water cluster for the HOMO and the LUMO for the different approximate strategies employed here as well as a combination thereof. Again, we find that the LT approximation 
does not introduce significant errors in QP energies for either the HOMOs or the LUMOs. For the NAF approximation ($\varepsilon_\text{NAF} = 10^{-2}$), the error with respect to the reference calculation is constant at about 1.5 meV and 3.0 meV for the HOMO and the LUMO, respectively. For the FC approximation, a constant error of about 3.5 meV and $-$0.5 meV is observable for the HOMO and the LUMO energies, respectively.
While the error of the approximation combining LT, NAF, and FC exceeds the individual errors in the HOMO case (about 4.5 meV), we find partial error cancellation in the LUMO case (about 1.8 meV). Most importantly, however, it can be seen that (a) the error in QP energies is essentially independent of the system size and (b) the magnitude of QP energy errors is within a tolerable range using the approximations and thresholds suggested here (compare Sec.~\ref{sec:compdetails}).

As a next step, we show computational timings of the various G$_0$W$_0$ methods.
We assess the practical scaling behavior with the system size by considering linear fits of double logarithmic plots of the wall-clock timings for the calculation of the screened Coulomb interaction $W_0$ [see, e.g., Eq.~(\ref{eq:ScreenedCoulombOrbital})] as a function of the number of SCF basis functions in Fig.~\ref{fig:doublelogplot}.
A non-logarithmic wall-clock timing plot along with the resulting speed-ups can be found in Fig.~S1 of the Supporting Information.
It can be argued that considering the slopes of the linear fits could be considered unsuitable here, as the formal scaling of each variant of our LT-GW approach remains unchanged with $\mathcal{O}(N^4)$. As a result, no difference in slopes is to be expected between the approaches in the limit where the respective algorithmic step with the highest-order scaling behavior dominates the computation time. We add those fits mainly to estimate the practical scaling implications for typical system sizes in molecular quantum-chemistry applications.

Taking a look at the corresponding linear fits performed on the data in Fig.~\ref{fig:doublelogplot}, we find a slope of 3.34 for the unmodified AC-G$_0$W$_0$ algorithm, which is only slightly smaller than the formal scaling exponent of four that would be expected for the AC approach.
The exponent is reduced by both the FC and NAF approximations to 3.30 and 3.13, respectively,
where no such reduction would be expected for the exponent but rather for the prefactor only.
Here, we note that the number of NAFs included in the calculations is on average 25--30\% lower than the number of original auxiliary basis functions. 
For the water cluster containing 100 water molecules, the auxiliary-basis size reduction is 26\%, which should result in a speed-up of $0.74^{-2} \approx 1.83$, and which is close to the observed speed-up of 2.0.
The LT approximation leads to a lowering of the exponent from 3.34 to 2.78.
In this case, the expected speed-up should be proportional to the quotient of the original number of imaginary frequencies and the number of Laplace grid points (see Eq.~\ref{eq:LT}).
For the cluster containing 100 water molecules, this ratio is $128/17 \approx 7.5$ which compares well with the observed speed-up of 6.7. Inspecting the exponents of the two combined approximations LT/NAF as well as LT/NAF/FC, we find that the individual reductions in computational scaling add up so that for LT/NAF/FC the slope of the linear fit (as a measure of the computational scaling) is lowered by almost one with respect to the regular AC-G$_0$W$_0$ calculation. For the presented wall-clock timings, it can thus be seen that, although the formal scaling behavior is unchanged by the approximations introduced, LT-G$_0$W$_0$ leads to a drastically lower practical computational scaling, because the onset of the asymptotic formal scaling of $\mathcal{O}(N^4)$ is delayed to larger systems than employed here, all while retaining a very high degree of accuracy. This makes LT-GW especially appealing for typical molecular quantum chemistry and also for subsystem DFT-based GW applications\cite{toelle2021}, where the fragments are typically chosen to be of medium size.
\begin{figure}[!t]
    \centering
    \includegraphics[width=0.7\linewidth]{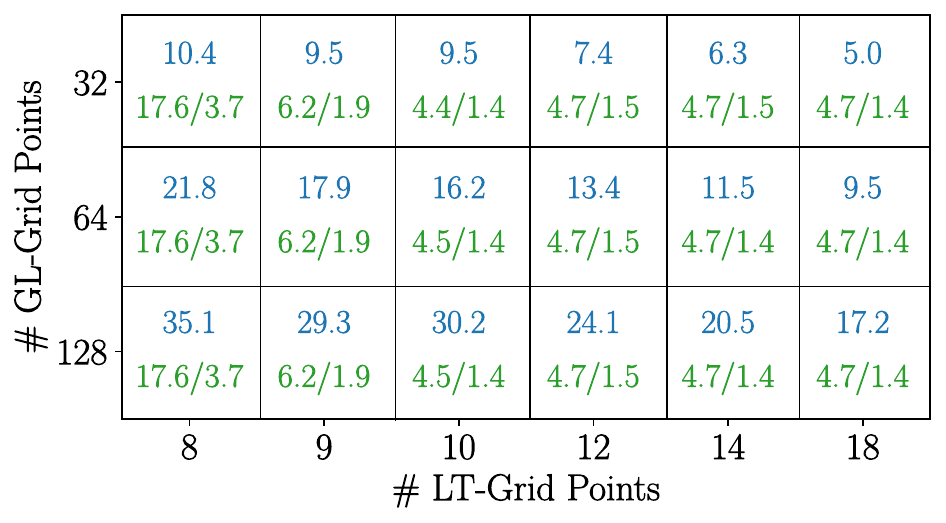}
    \caption{Comparison of the speed-up (blue) and deviation of the HOMO/LUMO quasi-particle energies (green, in eV), relative to AC-G$_0$W$_0$ for various Gauss--Legendre (GL) and LT grid sizes (in combination with FC/NAF) for a water cluster containing 100 water molecules.}
    \label{fig:GridComparison}
\end{figure}

As already indicated before, the thresholds for the numerical frequency integration and LT transformation are chosen conservatively. Therefore, a more in-depth comparison of accuracy and speed-up for various sizes of the modified GL-grid (32/64/128) and the LT-grid (8/9/10/12/14/18) in combination with the NAF and FC approximations are given in Fig.~\ref{fig:GridComparison}. 
The speed-ups range from $5.0$ to $35.1$ for a ratio of approximately 1.8 to 16 for the two grid sizes (resulting in a deviation in the HOMO/LUMO quasi-particle energies of $4.7/1.4$ and $17.6/3.7$~eV, respectively). 
This indicates again that, even though the ratio of the two grid sizes is small, a five-fold speed-up can be achieved in combination with NAF and FC because the total speed-up is the result of the product of the different contributions (LT/FC/NAF). 
For the remaining part of the manuscript we keep the default settings for the LT- and GL-grid as described in Sec.~\ref{sec:compdetails}. These settings introduce negligible errors, while resulting in realistic speed-ups which are neither at the low end nor at the high end of the spectrum, as can be deduced from Fig.~\ref{fig:GridComparison}.

Additionally, we consider absolute timings of the G$_0$W$_0$ and eigenvalue-self-consistent GW (five cycles) calculations for the cluster containing 100 water molecules to illustrate the speed-up that can be expected in practical calculations with moderately sized systems and the LT-G$_0$W$_0$ method. The results can be found in Tab.~\ref{tab:evgwtimings}.
\begin{table}[!t]
    \centering
    \caption{Wall-clock timings (min) and speed-ups for the calculation of the screened Coulomb interaction $W$ for G$_0$W$_0$ and eigenvalue-self-consistent GW (five cycles) calculations for the different approximations employed in this study. The largest water cluster consisting of 100 water molecules served as the test system.}
    \label{tab:evgwtimings}
    \begin{tabular}{lrrrrrr}
        \toprule
        & None & LT & NAF & FC & LT/NAF & LT/NAF/FC \\
        \midrule
        \multicolumn{7}{c}{ G$_0$W$_0$ }\\
        Wall-clock time (min)  & 711.0 & 105.7 & 350.4 & 570.4 & 49.7 & 39.3 \\
        Speed-up wrt reference &  1.0 & 6.7 & 2.0 & 1.2 & 14.3 & 18.1 \\
        \midrule
        \midrule
        \multicolumn{7}{c}{ evGW }\\
        Wall-clock time (min)  & 3502.6 & 530.0 & 1692.0 & 2775.3 & 263.2 & 199.4 \\
        Speed-up wrt reference & 1.0 & 6.6 & 2.1 & 1.3 & 13.3 & 17.6 \\
        \bottomrule
    \end{tabular}
\end{table}
It turns out that the speed-ups of the composite approximation LT/NAF/FC are 18.1 and 17.6 for G$_0$W$_0$ and evGW, respectively, which slightly exceeds the product of the speed-ups of the individual LT (6.7 and 6.6), NAF (2.0 and 2.1), and FC (1.2 and 1.3) approximations, each amounting to roughly 16. The individual approximations thus do not interfere with each other but can constructively be used in combination, and the respective speed-up directly carries over to (partially) self-consistent GW calculations. Additionally, in Fig. S2 of the SI, we break down the computational time as a function of the size of the water cluster both for regular and LT/NAF/FC-AC-G$_0$W$_0$ calculations into the contribution of (i) the three-center molecular-orbital integrals, (ii) the screened Coulomb interaction, and (iii) the NAF approximation (in the latter case). Further, we compare wall-clock timing contributions as a function of the employed CPU threads for the water cluster containing 100 water molecules in Tab.~S3 of the SI. Finally, we note that the G$_0$W$_0$ calculation using only the LT approximation is about twice as fast as the regular one already for the smallest investigated water cluster containing 10 molecules (10 seconds vs 20 seconds), providing evidence for the small prefactor of LT-GW combined with the NAF and FC approximations.

\subsection{LT-G$_0$W$_0$ with BSE}
We apply a combination of LT-G$_0$W$_0$ and the Bethe--Salpeter (BSE) equation to investigate the effect of the LT approximation on the accuracy of linear absorption spectra. The BSE calculations are performed with the efficient integral-direct resolution of the identity implementation for the Hartree--Fock and long-range exchange part of the response matrix in \textsc{Serenity} originally presented
in our work in Ref.~\citenum{hellmann2022automated}. 
As introduced above, the LT-G$_0$W$_0$ method refers to the application of the LT, NAF, and FC approximation and will be used in the following.
\begin{figure}[!t]
    \centering
    \includegraphics[width=0.95\linewidth]{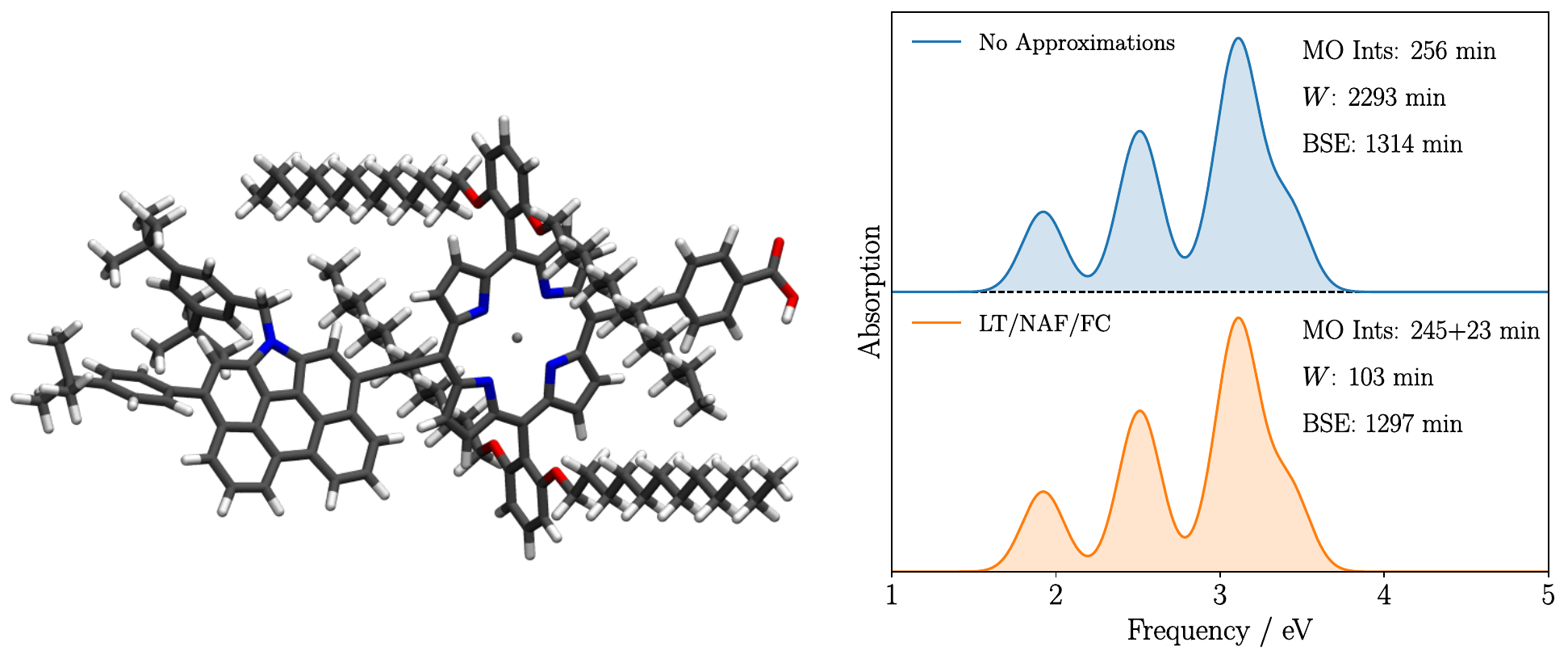}
    \caption{Comparison of linear absorption spectra obtained with a regular AC-G$_0$W$_0$/BSE calculation
    and one where the set of LT/NAF/FC approximations is used for the G$_0$W$_0$ calculation. The largest portions of the computational wall-clock timings are additionally broken down [BHLYP/def2-TZVP]. The wall-clock time for the reference KS-DFT calculation was 136 minutes. The oscillator strengths were calculated in the dipole-length representation and broadened with Gaussian functions with a full width at half maximum of 0.3 eV.}
    \label{fig:ww6spec}
\end{figure}

As a first test case, we consider the WW-6 dye relevant in photovoltaics~\cite{mester2019reduced,luo2014n}. The molecular geometry was taken from Ref.~\citenum{luo2014n} and is displayed in Fig.~\ref{fig:ww6spec}. Within the def2-TZVP basis set, there are 5583 SCF basis functions as well as 13802 auxiliary basis functions for the GW/BSE part of the calculation. In Fig.~\ref{fig:ww6spec}, we compare the linear absorption spectra for the WW-6 system that was obtained with the regular AC-G$_0$W$_0$/BSE calculation with the LT-G$_0$W$_0$ calculation employing both the NAF ($\varepsilon_\mathrm{NAF} = 10^{-2}$) and the FC approximations. 
In both cases, eight of the lowest-lying excitation energies and corresponding oscillator strengths were determined.
The FC approximation was not applied for the BSE calculations. We find no visible difference between the linear absorption spectra calculated with the regular and the approximate approach. Numerical results for QP energies as well as excitation energies and oscillator strengths can be found in Tabs.~\ref{tab:ww6errors1} and~\ref{tab:ww6errors2}, respectively. 
The mean deviation of QP energies is about 9.6 meV which far exceeds the mean error of 
excitation energies and oscillator strengths which amount to 0.75~meV and
$0.39\cdot 10^{-3}$~a.u., respectively. The occupied and virtual QP energy errors are more systematic for this test system than for the HOMOs and LUMOs of the water clusters investigated beforehand. This results in more favorable error cancellation for excitation energies, which depend on QP energy differences. The errors of the oscillator strengths are equally negligible, which, in turn, is probably a result of the eigenvectors of the BSE problem being largely unaffected because of the error cancellation mentioned above.
\begin{table}[!t]
    \centering
    \caption{Quasi-particle (QP) energies of the regular WW-6 G$_0$W$_0$ calculation ($\epsilon$),
    QP energies of the LT/NAF/FC-G$_0$W$_0$ calculation ($\tilde\epsilon$) as well as their deviation (BHLYP/def2-TZVP, MAE: mean-absolute error).}
    \label{tab:ww6errors1}
    \begin{tabular}{lrrr}
    \toprule
     & \phantom{12345}$\epsilon$ / eV & \phantom{12345}$\tilde\epsilon$ / eV & \phantom{123}$\Delta\epsilon$ / meV\\
    \midrule	
HOMO$-$4	&	$-$7.5346	&	$-$7.5262	&	8.3	\\
HOMO$-$3	&	$-$7.1752	&	$-$7.1662	&	9.0	\\
HOMO$-$2	&	$-$6.4024	&	$-$6.3932	&	9.2	\\
HOMO$-$1	&	$-$6.2786	&	$-$6.2691	&	9.5	\\
HOMO	&	$-$5.8659	&	$-$5.8564	&	9.5	\\
LUMO	&	$-$1.8539	&	$-$1.8438	&	10.0 \\
LUMO+1	&	$-$1.3771	&	$-$1.3670	&	10.1 \\
LUMO+2	&	$-$0.9054	&	$-$0.8951	&	10.3 \\
LUMO+3	&	$-$0.3816	&	$-$0.3717	&	 9.9 \\
LUMO+4	&	   0.1719	&	   0.1821	&	10.2 \\
\midrule
MAE     &               &               &    9.6 \\
\bottomrule
    \end{tabular}
\end{table}

Inspecting the computational timings (given in Fig.~\ref{fig:ww6spec}), we find that in the regular case, the overall wall-clock timings are dominated by the calculation of the screened Coulomb interaction $W$ with 2293 minutes, while in the approximate case, the BSE part of the calculation exceeds the time needed for the GW calculation by far.
Here, the overall G$_0$W$_0$ calculation time is, in fact, dominated by the preparation of the three-index MO integrals, as the calculation of $W$ only took 103 minutes.
We also note that for the approximate calculation, setting up the NAF matrix, diagonalizing it, and then performing the NAF transformation to the three-index integral tensor introduces a small overhead of about 25 minutes (or ten percent), which is summarized in the timings for the ``MO Ints''.
The number of NAFs included in the calculation was 8755 corresponding to a reduction of 37\% with respect to the full number of auxiliary basis functions. The speed-up for the entire calculation amounts to 2.3 (3915 minutes vs 1720 minutes) while the speed-up for the calculation of the screened Coulomb interaction alone is 22.3 (2293 minutes vs 103 minutes). These calculations demonstrate that LT-GW is able to provide accurate references for BSE calculations, while drastically reducing the computational demand of the preceding G$_0$W$_0$ calculation.
\begin{table}[!t]
    \centering
    \caption{Excitation energies ($\omega_{0n}$) and oscillator strengths ($f_{0n}$) of the 
    regular G$_0$W$_0$/BSE calculation and the LT/NAF/FC-G$_0$W$_0$/BSE calculation (indicated by a tilde) and the resulting deviation (BHLYP/def2-TZVP, MAE: mean-absolute error).}
    \label{tab:ww6errors2}
    \begin{tabular}{lrrrrrr}
    \toprule
$0\to n$ & $\omega_n$ / eV & $\tilde \omega_n$ / eV & $\Delta \omega_n$ / meV &
$f_n$ & $\tilde f_n$ & $\Delta f_n$ / $10^{-3}$ \\
\midrule
1	&	1.9220	&	1.9225	&	0.53	&	0.8508	&	0.8510	&	0.20	\\
2	&	1.9997	&	2.0002	&	0.52	&	0.0103	&	0.0103	&	0.01	\\
3	&	2.5107	&	2.5114	&	0.69	&	1.7237	&	1.7247	&	0.99	\\
4	&	2.9090	&	2.9098	&	0.77	&	0.1397	&	0.1396	&	$-$0.09	\\
5	&	3.0747	&	3.0754	&	0.62	&	0.5167	&	0.5172	&	0.47	\\
6	&	3.1149	&	3.1157	&	0.86	&	2.1007	&	2.1017	&	0.99	\\
7	&	3.2286	&	3.2297	&	1.07	&	0.0537	&	0.0536	&	$-$0.02	\\
8	&	3.4163	&	3.4172	&	0.91	&	0.9013	&	0.9017	&	0.37	\\
\midrule
MAE &           &           &   0.75    &           &           &   0.39    \\
\bottomrule
    \end{tabular}
\end{table}

\begin{figure}[!t]
    \centering
    \includegraphics[width=0.95\linewidth]{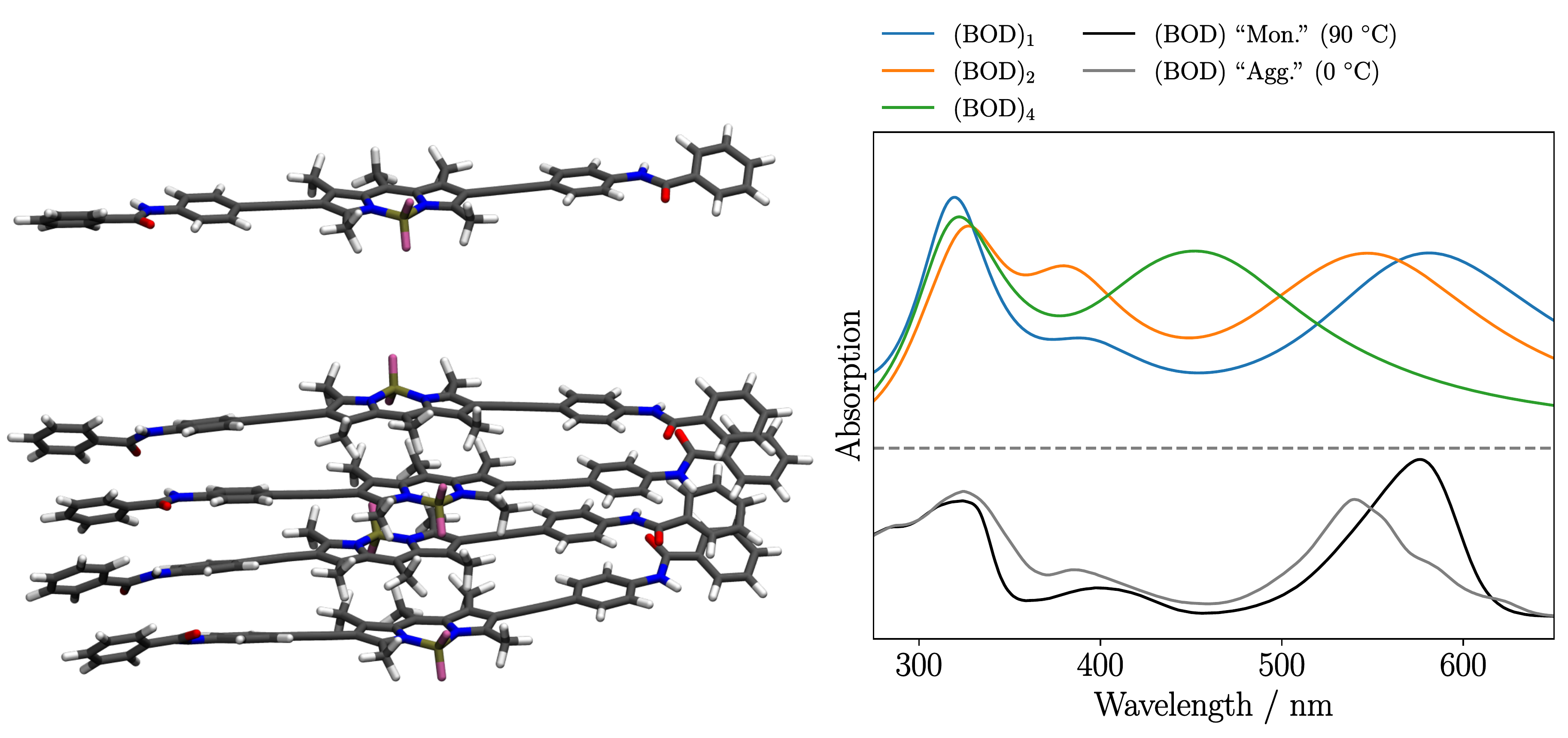}
    \caption{Left: BODIPY-based dyes in monomeric (upper) and tetrameric form (lower) (geometries provided by the authors of Ref.~{\protect\citenum{rodle2016influence}}). Right: Comparison of linear absorption spectra obtained with LT-GW [BHLYP/def2-TZVP/NAF/FC] with experimental absorption spectra originally presented in Ref.~{\protect\citenum{rodle2016influence}}, which were generated from raw data provided by the authors from that article. The experimental monomer (``Mon.'') and aggregate (``Agg.'') spectra were obtained by varying the solution temperature assuming that for very high and low temperatures, respectively, mainly the monomeric and aggregated form exist. The shown spectra correspond to the coldest and hottest solution temperatures investigated in Ref.~{\protect\citenum{rodle2016influence}}, namely 0 and 90~$^\circ$C. While methylcyclohexane was used as a solvent experimentally, we performed vacuum calculations. The oscillator strengths were calculated in the dipole-length representation and broadened with Gaussian functions with a full width at half maximum of 0.3 eV. The absorption spectra were subsequently red-shifted by 0.48 eV and converted to the wavelength domain.}
    \label{fig:bodipy}
\end{figure}

As a second test system, we consider stacks of BODIPY dyes, which are of interest in the field of supramolecular polymer design~\cite{rodle2016influence,aida2020supramolecular}. Additionally, supermolecular BODIPY-based compounds are interesting for GW/BSE calculations in particular because alternative (standard) methods for predicting their absorption spectra may either lack the necessary accuracy (e.g.~linear response time-dependent density-functional theory, see e.g. Ref.~\citenum{momeni2015td}) or are simply not feasible for this kind of system size (e.g.~coupled cluster-based methodology such as coupled cluster with singles and approximate doubles~\cite{christiansen1995} and even local variants thereof~\cite{berraud2019unveiling,feldt2021assessment}).
In our calculations, we include monomer, dimer, and tetramer geometries (provided by the authors of Ref.~\citenum{rodle2016influence} and displayed in Fig.~\ref{fig:bodipy}) and compare our G$_0$W$_0$/BSE-based spectra with experimental ones in Fig.~\ref{fig:bodipy}. For all $n$-mers, 32 of the lowest-lying excitation energies and corresponding oscillator strengths were determined after calculating 20 of both the lowest-lying virtual and highest-lying occupied QP energies for each monomer in each geometry, that is, 40 for the dimer as well as 80 for the tetramer. Based on the findings of the approximate calculations for the WW-6 test system, we omit G$_0$W$_0$ calculations that do not apply any further approximations here.

The experimental spectra exhibit three main bands at about 600, 400, and 300 nm. Interestingly, a strong blue shift of, in particular, the energetically lowest-lying absorption band is observed upon aggregation (experimentally induced by lowering the solution temperature). This behavior can most likely be attributed to the corresponding interaction of the transition dipole moments of the monomers in this stacking pattern. Going over to the computed spectra, one finds that the monomer spectrum reproduces the position and intensity of the experimental bands with a high degree of accuracy (given a constant shift of the absorption spectrum of 0.48 eV). It can further be seen that the blue shift of the lowest-lying absorption band of the dimer compares well with the experimental one. The computed tetramer spectrum exhibits a blue shift far exceeding the experimental one. This is most likely due to a combination of different factors. On the one hand, the experimental spectrum is a combination of several different aggregates of varying sizes and particular arrangements. On the other hand, the tetramer geometry was obtained by stacking two dimers on top of each other followed by a reoptimization. As a result, the distance between the inner two monomers is smaller than the distance between the outer pairs which could lead to an overestimation of the excitonic couplings leading to the blue shift.
The GW calculation (screened Coulomb interaction $W$) took 6, 70, and 813 minutes for the monomer, dimer, and tetramer, respectively.

\section{Conclusion}
We have presented the LT-GW method, for which we numerically demonstrated that it follows our three main objectives: (a) a small prefactor, (b) minimal effort for adaptation in existing AC-GW codes, and (c) significant performance
improvements (up to 22-fold) for a wide range of system sizes with controllable error.
For this, LT-GW combines the GW approximation in the context of the analytic continuation (AC) approach with a Laplace transformation (LT), natural auxiliary functions (NAFs), and the frozen-core (FC) approximation.
We have highlighted its synergy with the BSE for calculations of excitation energy and properties for extended systems consisting of up to 7412 basis functions.
We are convinced that the LT-GW method constitutes a practical and widely applicable extension to existing GW implementations for molecular systems.

In the LT-G$_0$W$_0$/BSE calculations, we have shown that the computational time is now dominated by the BSE calculation.
Based on our three guiding principles, we aim to achieve similar improvements also for the BSE in the future by making use of, for example, minimal auxiliary basis sets \cite{zhou2023minimal} or simplified integrals \cite{grimme2013simplified,cho2022simplified}.

\begin{suppinfo}
HOMO and LUMO quasi-particle energies for molecular systems of the GW100 benchmark set,
non-logarithmic wall-clock-timings and the speed-up plot of the water clusters,
wall-clock-timings contributions for the water clusters,
and wall-clock timings as a function of employed CPU threads for the water cluster containing 100 molecules
can be found in the Supporting Information.

The data supporting the findings of this study are available either within the supplementary material or upon reasonable request from the authors.
\end{suppinfo}

\begin{acknowledgement}
J.T. gratefully acknowledges funding by the Deutsche Forschungsgemeinschaft (DFG, German Research Foundation) through DFG-495279997.
N.N. and J.N. gratefully acknowledge funding by the DFG through SFB 1459 (Project A03, Project-ID A03-433682494).
We would like to thank Christian M\"uck-Lichtenfeld for providing the monomer, dimer, and tetramer BODIPY geometries originally presented in Ref.~\citenum{rodle2016influence}.
We would like to thank Alexander R\"odle and Gustavo Fern\'andez for providing the raw data of the experimental absorption spectra originally presented in Ref.~\citenum{rodle2016influence}.
\end{acknowledgement}

\bibliography{literatur}
\end{document}

\vspace*{0cm}
\begin{center}
{\LARGE
    Accelerating Analytic-Continuation GW Calculations with a Laplace Transformation and Natural Auxiliary Functions}
\vspace{2cm}

{\Large -- Supporting Information --}
\vspace{2cm}

{\large
Johannes T\"{o}lle$^{1,\ddagger}$\footnote{email: jtolle@caltech.edu},
Niklas Niemeyer$^{2,\ddagger}$, 
and Johannes Neugebauer$^2$\footnote{email: j.neugebauer@uni-muenster.de}}
\\[1cm]

$^1$Division of Chemistry and Chemical Engineering, \\
California Institute of Technology, Pasadena, California 91125, USA

$^2$University of M\"unster, Organisch-Chemisches Institut and\\
Center for Multiscale Theory and Computation,\\
Corrensstra{\ss}e 36, 48149 M\"unster, Germany
\vspace{2cm}

$^\ddagger$Both authors contributed equally.
\end{center}
\vfill

\begin{tabular}{ll}
Date: & \today \\
\end{tabular}
\thispagestyle{empty}

\clearpage

\section{GW100 Benchmark}

\begin{landscape} \tiny
\begin{longtable}{lrrrrrrrrrrr}
\caption{G$_0$W$_0$ HOMO quasi-particle energies for the molecular systems from the GW100 benchmark set for various approximations in the evaluation of the G$_0$W$_0$ self-energy relative to the ``fully-analytic'' approach without employing the RI approximation. Effective core potentials are used for the heavy elements rubidium, silver, xenon, and iodine. Analytic-RI: ``Fully-analytic'' approach using the RI approximation, AC: AC-G$_0$W$_0$, LT: AC-G$_0$W$_0$ in combination with AC-LT ($\varepsilon_\text{LT}=10^{-7}$), AC-FC: AC-G$_0$W$_0$ in combination with FC, AC-NAF: AC-G$_0$W$_0$ in combination with the NAF approximation ($\varepsilon_\text{NAF} = 10^{\{-6,-4,-2\}}$), AC-FC-LT-NAF: Combining AC-G$_0$W$_0$ with FC/LT/NAF ($\varepsilon_\text{LT}=10^{-7}$, $\varepsilon_\text{NAF} = 10^{\{-6,-4,-2\}}$) [def2-TZVP, starting from Hartree--Fock orbitals]. This data is additionally provided as a text file.} \label{homogw100}\\
\toprule
                       & Analytic RI & AC        & AC-FC     & AC-LT     & AC-NAF(6) & AC-NAF(4) & AC-NAF(2) & AC-LT-FC-NAF(6) & AC-LT-FC-NAF(4) & AC-LT-FC-NAF(2) & Analytic  \\
\midrule
        butane & -12.074 & -12.074 & -12.069 & -12.074 & -12.074 & -12.074 & -12.073 & -12.069 & -12.069 & -12.068 & -12.076 \\ \hline
        carbon tetrabromide & -10.728 & -10.728 & -10.808 & -10.728 & -10.728 & -10.728 & -10.727 & -10.808 & -10.808 & -10.807 & -10.726 \\ \hline
        hydrogen peroxide & -11.987 & -11.987 & -11.984 & -11.987 & -11.987 & -11.987 & -11.987 & -11.983 & -11.983 & -11.983 & -11.989 \\ \hline
        thymine & -9.596 & -9.596 & -9.589 & -9.596 & -9.596 & -9.596 & -9.595 & -9.589 & -9.589 & -9.588 & -9.596 \\ \hline
        fluorine & -16.265 & -16.265 & -16.264 & -16.265 & -16.265 & -16.265 & -16.265 & -16.264 & -16.264 & -16.263 & -16.266 \\ \hline
        toluene & -9.069 & -9.069 & -9.062 & -9.069 & -9.069 & -9.069 & -9.068 & -9.062 & -9.062 & -9.061 & -9.069 \\ \hline
        ethylbenzene & -9.038 & -9.038 & -9.031 & -9.038 & -9.038 & -9.038 & -9.037 & -9.031 & -9.031 & -9.030 & -9.038 \\ \hline
        arsenic dimer & -9.704 & -9.704 & -9.759 & -9.704 & -9.704 & -9.704 & -9.704 & -9.759 & -9.759 & -9.759 & -9.705 \\ \hline
        ethane & -13.036 & -13.036 & -13.033 & -13.036 & -13.036 & -13.036 & -13.033 & -13.033 & -13.033 & -13.030 & -13.037 \\ \hline
        benzene & -9.438 & -9.438 & -9.431 & -9.438 & -9.438 & -9.438 & -9.437 & -9.431 & -9.431 & -9.430 & -9.438 \\ \hline
        sulfur tetrafluoride & -13.214 & -13.214 & -13.222 & -13.214 & -13.214 & -13.214 & -13.214 & -13.222 & -13.222 & -13.221 & -13.213 \\ \hline
        tetracarbon & -11.557 & -11.557 & -11.554 & -11.557 & -11.557 & -11.557 & -11.556 & -11.554 & -11.554 & -11.553 & -11.557 \\ \hline
        fluoroborane & -11.264 & -11.264 & -11.260 & -11.264 & -11.264 & -11.264 & -11.263 & -11.260 & -11.260 & -11.259 & -11.264 \\ \hline
        potassium bromide & -8.176 & -8.176 & -8.241 & -8.176 & -8.176 & -8.176 & -8.175 & -8.241 & -8.241 & -8.239 & -8.176 \\ \hline
        cytosine & -9.199 & -9.199 & -9.193 & -9.199 & -9.199 & -9.199 & -9.199 & -9.193 & -9.193 & -9.192 & -9.200 \\ \hline
        krypton & -13.968 & -13.968 & -14.049 & -13.968 & -13.968 & -13.968 & -13.968 & -14.049 & -14.049 & -14.049 & -13.968 \\ \hline
        diborane(6) & -12.667 & -12.667 & -12.665 & -12.667 & -12.667 & -12.667 & -12.667 & -12.665 & -12.665 & -12.665 & -12.669 \\ \hline
        water & -12.778 & -12.778 & -12.775 & -12.778 & -12.778 & -12.778 & -12.777 & -12.774 & -12.774 & -12.773 & -12.780 \\ \hline
        formic acid & -11.878 & -11.878 & -11.875 & -11.878 & -11.878 & -11.878 & -11.877 & -11.875 & -11.875 & -11.874 & -11.879 \\ \hline
        propane & -12.484 & -12.484 & -12.481 & -12.484 & -12.484 & -12.484 & -12.482 & -12.481 & -12.481 & -12.479 & -12.486 \\ \hline
        phosphine & -10.683 & -10.683 & -10.691 & -10.683 & -10.683 & -10.683 & -10.681 & -10.691 & -10.690 & -10.689 & -10.685 \\ \hline
        aniline & -8.258 & -8.258 & -8.250 & -8.258 & -8.258 & -8.258 & -8.257 & -8.250 & -8.250 & -8.249 & -8.257 \\ \hline
        carbon monoxide & -15.003 & -15.003 & -14.999 & -15.003 & -15.003 & -15.003 & -15.002 & -14.999 & -14.999 & -14.998 & -15.004 \\ \hline
        lithium hydride & -7.944 & -7.944 & -7.944 & -7.944 & -7.944 & -7.944 & -7.940 & -7.944 & -7.944 & -7.940 & -7.946 \\ \hline
        sodium hexamer & -4.413 & -4.413 & -4.413 & -4.413 & -4.413 & -4.413 & -4.409 & -4.413 & -4.413 & -4.409 & -4.414 \\ \hline
        magnesium fluoride & -13.790 & -13.790 & -13.789 & -13.790 & -13.790 & -13.790 & -13.788 & -13.789 & -13.789 & -13.787 & -13.791 \\ \hline
        sulfur dioxide & -12.872 & -12.872 & -12.875 & -12.872 & -12.872 & -12.872 & -12.872 & -12.875 & -12.875 & -12.874 & -12.872 \\ \hline
        disilane & -10.978 & -10.978 & -10.995 & -10.978 & -10.978 & -10.978 & -10.977 & -10.995 & -10.995 & -10.993 & -10.978 \\ \hline
        copper dimer & -6.991 & -6.991 & -6.992 & -6.991 & -6.991 & -6.991 & -6.989 & -6.992 & -6.992 & -6.990 & -6.992 \\ \hline
        ozone & -10.953 & -10.953 & -10.949 & -10.953 & -10.953 & -10.953 & -10.952 & -10.949 & -10.949 & -10.947 & -10.955 \\ \hline
        nitrogen & -17.072 & -17.072 & -17.067 & -17.072 & -17.072 & -17.072 & -17.072 & -17.067 & -17.067 & -17.066 & -17.074 \\ \hline
        ammonia & -11.089 & -11.089 & -11.085 & -11.089 & -11.089 & -11.089 & -11.088 & -11.085 & -11.085 & -11.084 & -11.088 \\ \hline
        acetaldehyde & -10.722 & -10.722 & -10.719 & -10.722 & -10.722 & -10.722 & -10.721 & -10.719 & -10.719 & -10.717 & -10.724 \\ \hline
        hydrogen sulfide & -10.413 & -10.413 & -10.426 & -10.413 & -10.413 & -10.413 & -10.412 & -10.426 & -10.426 & -10.425 & -10.415 \\ \hline
        sodium chloride & -9.156 & -9.156 & -9.173 & -9.156 & -9.156 & -9.156 & -9.155 & -9.173 & -9.173 & -9.170 & -9.157 \\ \hline
        phosphorus dimer & -10.473 & -10.473 & -10.482 & -10.473 & -10.473 & -10.473 & -10.472 & -10.482 & -10.482 & -10.481 & -10.475 \\ \hline
        ethylene & -10.685 & -10.685 & -10.678 & -10.685 & -10.685 & -10.685 & -10.684 & -10.678 & -10.678 & -10.677 & -10.686 \\ \hline
        methane & -14.633 & -14.633 & -14.630 & -14.633 & -14.633 & -14.633 & -14.631 & -14.630 & -14.630 & -14.627 & -14.634 \\ \hline
        cyclopentadiene & -8.777 & -8.777 & -8.770 & -8.777 & -8.777 & -8.777 & -8.776 & -8.770 & -8.770 & -8.768 & -8.778 \\ \hline
        hydrazine & -10.068 & -10.068 & -10.064 & -10.068 & -10.068 & -10.068 & -10.067 & -10.064 & -10.064 & -10.063 & -10.067 \\ \hline
        cyclopropane & -11.199 & -11.199 & -11.192 & -11.199 & -11.199 & -11.199 & -11.199 & -11.192 & -11.192 & -11.191 & -11.201 \\ \hline
        vinyl bromide & -9.997 & -9.997 & -10.027 & -9.997 & -9.997 & -9.997 & -9.996 & -10.027 & -10.027 & -10.026 & -9.997 \\ \hline
        potassium hydride & -6.099 & -6.099 & -6.100 & -6.099 & -6.099 & -6.099 & -6.095 & -6.100 & -6.100 & -6.096 & -6.100 \\ \hline
        silane & -13.078 & -13.078 & -13.092 & -13.078 & -13.078 & -13.078 & -13.077 & -13.092 & -13.092 & -13.091 & -13.079 \\ \hline
        guanine & -8.355 & -8.355 & -8.347 & -8.355 & -8.355 & -8.355 & -8.354 & -8.347 & -8.347 & -8.346 & -8.356 \\ \hline
        neon & -21.349 & -21.349 & -21.351 & -21.349 & -21.349 & -21.349 & -21.349 & -21.350 & -21.350 & -21.350 & -21.350 \\ \hline
        hydrogen cyanide & -13.821 & -13.821 & -13.814 & -13.821 & -13.821 & -13.821 & -13.820 & -13.814 & -13.814 & -13.814 & -13.822 \\ \hline
        boron nitride & -11.692 & -11.692 & -11.685 & -11.692 & -11.692 & -11.692 & -11.690 & -11.685 & -11.685 & -11.683 & -11.692 \\ \hline
        urea & -10.604 & -10.604 & -10.599 & -10.604 & -10.604 & -10.604 & -10.603 & -10.599 & -10.599 & -10.598 & -10.604 \\ \hline
        lithium fluoride & -11.251 & -11.251 & -11.250 & -11.251 & -11.251 & -11.251 & -11.248 & -11.250 & -11.250 & -11.246 & -11.252 \\ \hline
        helium & -24.294 & -24.294 & -24.294 & -24.294 & -24.294 & -24.294 & -24.294 & -24.294 & -24.294 & -24.294 & -24.294 \\ \hline
        acetylene & -11.530 & -11.530 & -11.524 & -11.530 & -11.530 & -11.530 & -11.529 & -11.524 & -11.524 & -11.523 & -11.532 \\ \hline
        ethoxy ethane & -10.377 & -10.377 & -10.374 & -10.377 & -10.377 & -10.377 & -10.375 & -10.374 & -10.374 & -10.372 & -10.377 \\ \hline
        hydrogen fluoride & -16.148 & -16.148 & -16.146 & -16.148 & -16.148 & -16.148 & -16.147 & -16.146 & -16.146 & -16.146 & -16.149 \\ \hline
        cyclooctatetraene & -8.588 & -8.588 & -8.581 & -8.588 & -8.588 & -8.588 & -8.587 & -8.581 & -8.581 & -8.580 & -8.587 \\ \hline
        pyridine & -9.830 & -9.830 & -9.823 & -9.830 & -9.830 & -9.830 & -9.829 & -9.823 & -9.823 & -9.822 & -9.830 \\ \hline
        carbon tetrachloride & -11.913 & -11.913 & -11.934 & -11.913 & -11.913 & -11.913 & -11.913 & -11.934 & -11.934 & -11.932 & -11.913 \\ \hline
        pentasilane & -9.705 & -9.705 & -9.722 & -9.705 & -9.705 & -9.705 & -9.704 & -9.722 & -9.722 & -9.720 & -9.703 \\ \hline
        uracil & -10.009 & -10.009 & -10.003 & -10.009 & -10.009 & -10.009 & -10.009 & -10.003 & -10.003 & -10.002 & -10.009 \\ \hline
        methanol & -11.460 & -11.460 & -11.457 & -11.460 & -11.460 & -11.460 & -11.459 & -11.457 & -11.457 & -11.455 & -11.461 \\ \hline
        borane & -13.528 & -13.528 & -13.526 & -13.528 & -13.528 & -13.528 & -13.528 & -13.526 & -13.526 & -13.525 & -13.530 \\ \hline
        carbon tetrafluoride & -16.792 & -16.792 & -16.791 & -16.792 & -16.792 & -16.792 & -16.792 & -16.791 & -16.791 & -16.790 & -16.793 \\ \hline
        vinyl chloride & -10.285 & -10.285 & -10.285 & -10.285 & -10.285 & -10.285 & -10.284 & -10.285 & -10.285 & -10.284 & -10.286 \\ \hline
        phenol & -8.823 & -8.823 & -8.815 & -8.823 & -8.823 & -8.823 & -8.822 & -8.815 & -8.815 & -8.814 & -8.823 \\ \hline
        aluminum fluoride & -15.577 & -15.577 & -15.588 & -15.577 & -15.577 & -15.577 & -15.576 & -15.588 & -15.588 & -15.587 & -15.578 \\ \hline
        carbon disulfide & -10.238 & -10.238 & -10.251 & -10.238 & -10.238 & -10.238 & -10.238 & -10.251 & -10.251 & -10.250 & -10.239 \\ \hline
        lithium dimer & -5.157 & -5.157 & -5.157 & -5.157 & -5.157 & -5.157 & -5.156 & -5.157 & -5.157 & -5.156 & -5.160 \\ \hline
        carbon dioxide & -14.164 & -14.164 & -14.160 & -14.164 & -14.164 & -14.164 & -14.163 & -14.160 & -14.160 & -14.159 & -14.165 \\ \hline
        phosphorus mononitride & -12.303 & -12.303 & -12.305 & -12.303 & -12.303 & -12.303 & -12.302 & -12.305 & -12.305 & -12.304 & -12.305 \\ \hline
        argon & -15.682 & -15.682 & -15.706 & -15.682 & -15.682 & -15.682 & -15.682 & -15.706 & -15.706 & -15.704 & -15.683 \\ \hline
        hydrogen azide & -11.029 & -11.029 & -11.024 & -11.029 & -11.029 & -11.029 & -11.029 & -11.024 & -11.024 & -11.023 & -11.029 \\ \hline
        hexafluorobenzene & -10.557 & -10.557 & -10.549 & -10.557 & -10.557 & -10.557 & -10.556 & -10.549 & -10.549 & -10.548 & -10.558 \\ \hline
        vinyl fluoride & -10.750 & -10.750 & -10.743 & -10.750 & -10.750 & -10.750 & -10.749 & -10.743 & -10.743 & -10.742 & -10.751 \\ \hline
        beryllium monoxide & -9.761 & -9.761 & -9.758 & -9.761 & -9.761 & -9.761 & -9.759 & -9.758 & -9.758 & -9.757 & -9.762 \\ \hline
        carbon oxysulfide & -11.474 & -11.474 & -11.483 & -11.474 & -11.474 & -11.474 & -11.473 & -11.483 & -11.483 & -11.482 & -11.475 \\ \hline
        formaldehyde & -11.268 & -11.268 & -11.265 & -11.268 & -11.268 & -11.268 & -11.267 & -11.265 & -11.265 & -11.263 & -11.269 \\ \hline
        ethanol & -11.172 & -11.172 & -11.169 & -11.172 & -11.172 & -11.172 & -11.171 & -11.169 & -11.169 & -11.167 & -11.173 \\ \hline
        bromine & -10.714 & -10.714 & -10.793 & -10.714 & -10.714 & -10.714 & -10.714 & -10.793 & -10.793 & -10.792 & -10.712 \\ \hline
        germane & -12.732 & -12.732 & -12.766 & -12.732 & -12.732 & -12.732 & -12.732 & -12.766 & -12.766 & -12.765 & -12.731 \\ \hline
        magnesium chloride & -11.851 & -11.851 & -11.868 & -11.851 & -11.851 & -11.851 & -11.850 & -11.868 & -11.868 & -11.866 & -11.852 \\ \hline
        carbon oxyselenide & -10.653 & -10.653 & -10.705 & -10.653 & -10.653 & -10.653 & -10.652 & -10.705 & -10.705 & -10.705 & -10.653 \\ \hline
        titanium fluoride & -16.048 & -16.048 & -16.047 & -16.048 & -16.048 & -16.048 & -16.048 & -16.047 & -16.047 & -16.046 & -16.048 \\ \hline
        copper cyanide & -11.187 & -11.187 & -11.183 & -11.187 & -11.187 & -11.187 & -11.186 & -11.183 & -11.183 & -11.183 & -11.188 \\ \hline
        sodium tetramer & -4.241 & -4.241 & -4.241 & -4.241 & -4.241 & -4.241 & -4.237 & -4.241 & -4.241 & -4.237 & -4.243 \\ \hline
        hydrogen chloride & -12.710 & -12.710 & -12.728 & -12.710 & -12.710 & -12.710 & -12.710 & -12.728 & -12.728 & -12.727 & -12.712 \\ \hline
        arsine & -10.495 & -10.495 & -10.553 & -10.495 & -10.495 & -10.495 & -10.494 & -10.553 & -10.553 & -10.551 & -10.499 \\ \hline
        magnesium monoxide & -8.383 & -8.383 & -8.382 & -8.383 & -8.383 & -8.383 & -8.380 & -8.382 & -8.382 & -8.379 & -8.384 \\ \hline
        dipotassium & -4.025 & -4.025 & -4.025 & -4.025 & -4.025 & -4.025 & -4.020 & -4.025 & -4.025 & -4.020 & -4.026 \\ \hline
        sodium dimer & -4.933 & -4.933 & -4.933 & -4.933 & -4.933 & -4.933 & -4.928 & -4.933 & -4.933 & -4.928 & -4.930 \\ \hline
        chlorine & -11.686 & -11.686 & -11.707 & -11.686 & -11.686 & -11.686 & -11.686 & -11.707 & -11.707 & -11.705 & -11.686 \\ \hline
        adenine & -8.624 & -8.624 & -8.616 & -8.624 & -8.624 & -8.624 & -8.623 & -8.616 & -8.616 & -8.615 & -8.624 \\ \hline
        gallium monochloride & -9.856 & -9.856 & -9.909 & -9.856 & -9.856 & -9.856 & -9.855 & -9.909 & -9.909 & -9.909 & -9.854 \\ \hline
        hydrogen & -16.304 & -16.304 & -16.304 & -16.304 & -16.304 & -16.304 & -16.304 & -16.304 & -16.304 & -16.304 & -16.306 \\ \hline
        dirubidium & -3.837 & -3.837 & -3.837 & -3.837 & -3.837 & -3.837 & -3.833 & -3.837 & -3.837 & -3.833 & -3.844 \\ \hline
        silver dimer & -6.963 & -6.963 & -6.963 & -6.963 & -6.963 & -6.963 & -6.960 & -6.963 & -6.963 & -6.960 & -6.964 \\ \hline
        xenon & -12.324 & -12.324 & -12.371 & -12.324 & -12.324 & -12.324 & -12.324 & -12.371 & -12.371 & -12.371 & -12.324 \\ \hline
        iodine & -9.674 & -9.674 & -9.717 & -9.674 & -9.674 & -9.674 & -9.673 & -9.717 & -9.717 & -9.717 & -9.672 \\ \hline
        vinyl iodide & -9.456 & -9.456 & -9.481 & -9.456 & -9.456 & -9.456 & -9.455 & -9.481 & -9.481 & -9.481 & -9.456 \\ \hline
        aluminum iodide & -9.970 & -9.970 & -10.013 & -9.970 & -9.970 & -9.970 & -9.968 & -10.013 & -10.013 & -10.013 & -9.969 \\ \hline
        carbon tetraiodide & -9.500 & -9.500 & -9.545 & -9.500 & -9.500 & -9.500 & -9.499 & -9.545 & -9.545 & -9.545 & -9.499 \\ 
\bottomrule
\end{longtable}

\clearpage

\begin{longtable}{lrrrrrrrrrrr}
\caption{G$_0$W$_0$ LUMO quasi-particle energies for the molecular systems from the GW100 benchmark set for various approximations in the evaluation of the G$_0$W$_0$ self-energy relative to the ``fully-analytic'' approach without employing the RI approximation. Effective core potentials are used for the heavy elements rubidium, silver, xenon, and iodine. Analytic-RI: ``Fully-analytic'' approach using the RI approximation, AC: AC-G$_0$W$_0$, LT: AC-G$_0$W$_0$ in combination with AC-LT ($\varepsilon_\text{LT}=10^{-7}$), AC-FC: AC-G$_0$W$_0$ in combination with FC, AC-NAF: AC-G$_0$W$_0$ in combination with the NAF approximation ($\varepsilon_\text{NAF} = 10^{\{-6,-4,-2\}}$), AC-FC-LT-NAF: Combining AC-G$_0$W$_0$ with FC/LT/NAF ($\varepsilon_\text{LT}=10^{-7}$, $\varepsilon_\text{NAF} = 10^{\{-6,-4,-2\}}$) [def2-TZVP, starting from Hartree--Fock orbitals]. This data is additionally provided as a text file.} \label{lumogw100} \\
\toprule
                       & Analytic RI & AC        & AC-FC     & AC-LT     & AC-NAF(6) & AC-NAF(4) & AC-NAF(2) & AC-LT-FC-NAF(6) & AC-LT-FC-NAF(4) & AC-LT-FC-NAF(2) & Analytic  \\
\midrule
butane & 3.131 & 3.131 & 3.132 & 3.131 & 3.131 & 3.131 & 3.136 & 3.132 & 3.132 & 3.140 & 3.126 \\ \hline
        carbon tetrabromide & -0.254 & -0.254 & -0.293 & -0.254 & -0.254 & -0.254 & -0.253 & -0.293 & -0.293 & -0.292 & -0.256 \\ \hline
        hydrogen peroxide & 3.269 & 3.269 & 3.269 & 3.269 & 3.269 & 3.269 & 3.272 & 3.269 & 3.269 & 3.271 & 3.269 \\ \hline
        thymine & 0.848 & 0.848 & 0.855 & 0.848 & 0.848 & 0.848 & 0.849 & 0.855 & 0.855 & 0.857 & 0.848 \\ \hline
        fluorine & 0.814 & 0.814 & 0.815 & 0.814 & 0.814 & 0.814 & 0.814 & 0.815 & 0.815 & 0.816 & 0.809 \\ \hline
        toluene & 1.845 & 1.845 & 1.852 & 1.845 & 1.845 & 1.845 & 1.847 & 1.852 & 1.852 & 1.854 & 1.847 \\ \hline
        ethylbenzene & 2.002 & 2.002 & 2.008 & 2.002 & 2.002 & 2.002 & 2.005 & 2.008 & 2.008 & 2.011 & 2.006 \\ \hline
        arsenic dimer & -0.361 & -0.361 & -0.413 & -0.361 & -0.361 & -0.361 & -0.360 & -0.413 & -0.413 & -0.412 & -0.362 \\ \hline
        ethane & 3.342 & 3.342 & 3.342 & 3.342 & 3.342 & 3.342 & 3.347 & 3.342 & 3.342 & 3.349 & 3.340 \\ \hline
        benzene & 1.892 & 1.892 & 1.899 & 1.892 & 1.892 & 1.892 & 1.894 & 1.899 & 1.899 & 1.901 & 1.895 \\ \hline
        sulfur tetrafluoride & 1.035 & 1.035 & 1.034 & 1.035 & 1.035 & 1.035 & 1.036 & 1.034 & 1.034 & 1.035 & 1.031 \\ \hline
        tetracarbon & -2.193 & -2.193 & -2.187 & -2.193 & -2.193 & -2.193 & -2.192 & -2.187 & -2.187 & -2.184 & -2.193 \\ \hline
        fluoroborane & 1.641 & 1.641 & 1.643 & 1.641 & 1.641 & 1.641 & 1.642 & 1.643 & 1.643 & 1.645 & 1.641 \\ \hline
        potassium bromide & -0.382 & -0.382 & -0.383 & -0.382 & -0.382 & -0.382 & -0.372 & -0.383 & -0.382 & -0.374 & -0.382 \\ \hline
        cytosine & 1.014 & 1.014 & 1.021 & 1.014 & 1.014 & 1.014 & 1.016 & 1.021 & 1.021 & 1.024 & 1.015 \\ \hline
        krypton & 10.490 & 10.490 & 10.466 & 10.490 & 10.490 & 10.490 & 10.491 & 10.466 & 10.466 & 10.467 & 10.489 \\ \hline
        diborane(6) & 1.583 & 1.583 & 1.588 & 1.583 & 1.583 & 1.583 & 1.586 & 1.588 & 1.588 & 1.591 & 1.583 \\ \hline
        water & 3.126 & 3.126 & 3.125 & 3.126 & 3.126 & 3.126 & 3.126 & 3.125 & 3.125 & 3.126 & 3.125 \\ \hline
        formic acid & 3.320 & 3.320 & 3.320 & 3.320 & 3.320 & 3.320 & 3.324 & 3.320 & 3.320 & 3.323 & 3.320 \\ \hline
        propane & 3.184 & 3.184 & 3.185 & 3.184 & 3.184 & 3.184 & 3.189 & 3.185 & 3.185 & 3.192 & 3.180 \\ \hline
        phosphine & 3.317 & 3.317 & 3.320 & 3.317 & 3.317 & 3.317 & 3.318 & 3.320 & 3.320 & 3.321 & 3.323 \\ \hline
        aniline & 1.947 & 1.947 & 1.954 & 1.947 & 1.947 & 1.947 & 1.949 & 1.954 & 1.954 & 1.956 & 1.949 \\ \hline
        carbon monoxide & 1.150 & 1.150 & 1.155 & 1.150 & 1.150 & 1.150 & 1.151 & 1.155 & 1.155 & 1.157 & 1.151 \\ \hline
        lithium hydride & 0.128 & 0.128 & 0.128 & 0.128 & 0.128 & 0.128 & 0.127 & 0.128 & 0.128 & 0.127 & 0.125 \\ \hline
        sodium hexamer & -0.413 & -0.413 & -0.413 & -0.413 & -0.413 & -0.413 & -0.404 & -0.413 & -0.413 & -0.404 & -0.421 \\ \hline
        magnesium fluoride & 0.045 & 0.045 & 0.045 & 0.045 & 0.045 & 0.045 & 0.054 & 0.045 & 0.045 & 0.055 & 0.046 \\ \hline
        sulfur dioxide & -0.472 & -0.472 & -0.473 & -0.472 & -0.472 & -0.472 & -0.471 & -0.473 & -0.473 & -0.472 & -0.473 \\ \hline
        disilane & 2.599 & 2.599 & 2.606 & 2.599 & 2.599 & 2.599 & 2.603 & 2.606 & 2.606 & 2.609 & 2.595 \\ \hline
        copper dimer & -0.031 & -0.031 & -0.031 & -0.031 & -0.031 & -0.031 & -0.021 & -0.031 & -0.031 & -0.022 & -0.041 \\ \hline
        ozone & -4.108 & -4.108 & -4.104 & -4.108 & -4.108 & -4.108 & -4.107 & -4.104 & -4.104 & -4.102 & -4.108 \\ \hline
        nitrogen & 3.075 & 3.075 & 3.080 & 3.075 & 3.075 & 3.075 & 3.077 & 3.080 & 3.080 & 3.082 & 3.075 \\ \hline
        ammonia & 3.118 & 3.118 & 3.118 & 3.118 & 3.118 & 3.118 & 3.121 & 3.118 & 3.118 & 3.119 & 3.115 \\ \hline
        acetaldehyde & 2.181 & 2.181 & 2.186 & 2.181 & 2.181 & 2.181 & 2.183 & 2.186 & 2.186 & 2.188 & 2.183 \\ \hline
        hydrogen sulfide & 3.117 & 3.117 & 3.117 & 3.117 & 3.117 & 3.117 & 3.119 & 3.117 & 3.117 & 3.118 & 3.119 \\ \hline
        sodium chloride & -0.560 & -0.560 & -0.560 & -0.560 & -0.560 & -0.560 & -0.548 & -0.560 & -0.560 & -0.549 & -0.560 \\ \hline
        phosphorus dimer & -0.200 & -0.200 & -0.207 & -0.200 & -0.200 & -0.200 & -0.199 & -0.207 & -0.207 & -0.206 & -0.202 \\ \hline
        ethylene & 2.850 & 2.850 & 2.855 & 2.850 & 2.850 & 2.850 & 2.853 & 2.855 & 2.855 & 2.857 & 2.850 \\ \hline
        methane & 3.664 & 3.664 & 3.664 & 3.664 & 3.664 & 3.664 & 3.667 & 3.664 & 3.664 & 3.670 & 3.662 \\ \hline
        cyclopentadiene & 1.908 & 1.908 & 1.914 & 1.908 & 1.908 & 1.908 & 1.911 & 1.914 & 1.914 & 1.916 & 1.908 \\ \hline
        hydrazine & 2.799 & 2.799 & 2.800 & 2.799 & 2.799 & 2.799 & 2.803 & 2.800 & 2.800 & 2.802 & 2.792 \\ \hline
        cyclopropane & 3.725 & 3.725 & 3.725 & 3.725 & 3.725 & 3.725 & 3.729 & 3.725 & 3.725 & 3.732 & 3.725 \\ \hline
        vinyl bromide & 2.125 & 2.125 & 2.128 & 2.125 & 2.125 & 2.125 & 2.128 & 2.128 & 2.128 & 2.131 & 2.126 \\ \hline
        potassium hydride & -0.018 & -0.018 & -0.018 & -0.018 & -0.018 & -0.018 & -0.002 & -0.018 & -0.018 & -0.002 & -0.019 \\ \hline
        silane & 3.375 & 3.375 & 3.378 & 3.375 & 3.375 & 3.375 & 3.378 & 3.378 & 3.378 & 3.382 & 3.376 \\ \hline
        guanine & 2.116 & 2.116 & 2.117 & 2.116 & 2.116 & 2.116 & 2.122 & 2.117 & 2.117 & 2.121 & 2.112 \\ \hline
        neon & 21.198 & 21.198 & 21.195 & 21.198 & 21.198 & 21.198 & 21.197 & 21.195 & 21.194 & 21.195 & 21.199 \\ \hline
        hydrogen cyanide & 3.542 & 3.542 & 3.543 & 3.542 & 3.542 & 3.542 & 3.547 & 3.543 & 3.543 & 3.548 & 3.541 \\ \hline
        boron nitride & -3.830 & -3.830 & -3.823 & -3.830 & -3.830 & -3.830 & -3.829 & -3.823 & -3.823 & -3.821 & -3.832 \\ \hline
        urea & 2.570 & 2.570 & 2.569 & 2.570 & 2.570 & 2.570 & 2.575 & 2.569 & 2.569 & 2.573 & 2.565 \\ \hline
        lithium fluoride & -0.012 & -0.012 & -0.012 & -0.012 & -0.012 & -0.012 & -0.004 & -0.012 & -0.012 & -0.004 & -0.012 \\ \hline
        helium & 22.402 & 22.402 & 22.402 & 22.402 & 22.402 & 22.402 & 22.402 & 22.402 & 22.402 & 22.402 & 22.401 \\ \hline
        acetylene & 3.765 & 3.765 & 3.765 & 3.765 & 3.765 & 3.765 & 3.769 & 3.765 & 3.765 & 3.768 & 3.763 \\ \hline
        ethoxy ethane & 3.210 & 3.210 & 3.211 & 3.210 & 3.210 & 3.210 & 3.215 & 3.211 & 3.211 & 3.219 & 3.207 \\ \hline
        hydrogen fluoride & 3.232 & 3.232 & 3.232 & 3.232 & 3.232 & 3.232 & 3.236 & 3.232 & 3.232 & 3.236 & 3.233 \\ \hline
        cyclooctatetraene & 0.941 & 0.941 & 0.948 & 0.941 & 0.941 & 0.941 & 0.943 & 0.948 & 0.948 & 0.950 & 0.943 \\ \hline
        pyridine & 1.337 & 1.337 & 1.344 & 1.337 & 1.337 & 1.337 & 1.339 & 1.344 & 1.344 & 1.346 & 1.339 \\ \hline
        carbon tetrachloride & 1.112 & 1.112 & 1.106 & 1.112 & 1.112 & 1.112 & 1.113 & 1.106 & 1.106 & 1.106 & 1.112 \\ \hline
        pentasilane & 1.128 & 1.128 & 1.137 & 1.128 & 1.128 & 1.128 & 1.131 & 1.137 & 1.137 & 1.140 & 1.115 \\ \hline
        uracil & 0.771 & 0.771 & 0.778 & 0.771 & 0.771 & 0.771 & 0.772 & 0.778 & 0.778 & 0.780 & 0.772 \\ \hline
        methanol & 3.291 & 3.291 & 3.291 & 3.291 & 3.291 & 3.291 & 3.294 & 3.291 & 3.291 & 3.294 & 3.289 \\ \hline
        borane & 0.708 & 0.708 & 0.712 & 0.708 & 0.708 & 0.708 & 0.710 & 0.712 & 0.712 & 0.714 & 0.708 \\ \hline
        carbon tetrafluoride & 5.158 & 5.158 & 5.157 & 5.158 & 5.158 & 5.158 & 5.162 & 5.157 & 5.157 & 5.161 & 5.156 \\ \hline
        vinyl chloride & 2.292 & 2.292 & 2.297 & 2.292 & 2.292 & 2.292 & 2.295 & 2.297 & 2.297 & 2.299 & 2.293 \\ \hline
        phenol & 1.773 & 1.773 & 1.779 & 1.773 & 1.773 & 1.773 & 1.774 & 1.779 & 1.779 & 1.782 & 1.775 \\ \hline
        aluminum fluoride & 0.794 & 0.794 & 0.815 & 0.794 & 0.794 & 0.794 & 0.800 & 0.815 & 0.815 & 0.821 & 0.794 \\ \hline
        carbon disulfide & 0.204 & 0.204 & 0.204 & 0.204 & 0.204 & 0.204 & 0.205 & 0.204 & 0.204 & 0.204 & 0.204 \\ \hline
        lithium dimer & 0.039 & 0.039 & 0.039 & 0.039 & 0.039 & 0.039 & 0.042 & 0.039 & 0.039 & 0.042 & 0.028 \\ \hline
        carbon dioxide & 2.980 & 2.980 & 2.980 & 2.980 & 2.980 & 2.980 & 2.985 & 2.980 & 2.980 & 2.983 & 2.981 \\ \hline
        phosphorus mononitride & 0.354 & 0.354 & 0.353 & 0.354 & 0.354 & 0.354 & 0.355 & 0.353 & 0.353 & 0.354 & 0.354 \\ \hline
        argon & 14.838 & 14.838 & 14.833 & 14.838 & 14.838 & 14.838 & 14.838 & 14.833 & 14.833 & 14.833 & 14.837 \\ \hline
        hydrogen azide & 2.081 & 2.081 & 2.085 & 2.081 & 2.081 & 2.081 & 2.083 & 2.085 & 2.085 & 2.087 & 2.081 \\ \hline
        hexafluorobenzene & 0.910 & 0.910 & 0.920 & 0.910 & 0.910 & 0.910 & 0.911 & 0.920 & 0.920 & 0.922 & 0.910 \\ \hline
        vinyl fluoride & 2.957 & 2.957 & 2.963 & 2.957 & 2.957 & 2.957 & 2.961 & 2.963 & 2.963 & 2.965 & 2.958 \\ \hline
        beryllium monoxide & -2.086 & -2.086 & -2.086 & -2.086 & -2.086 & -2.086 & -2.075 & -2.086 & -2.086 & -2.077 & -2.088 \\ \hline
        carbon oxysulfide & 1.784 & 1.784 & 1.788 & 1.784 & 1.784 & 1.784 & 1.785 & 1.788 & 1.788 & 1.789 & 1.784 \\ \hline
        formaldehyde & 1.902 & 1.902 & 1.906 & 1.902 & 1.902 & 1.902 & 1.904 & 1.906 & 1.906 & 1.908 & 1.903 \\ \hline
        ethanol & 3.122 & 3.122 & 3.123 & 3.122 & 3.122 & 3.122 & 3.126 & 3.123 & 3.123 & 3.128 & 3.120 \\ \hline
        bromine & -0.747 & -0.747 & -0.818 & -0.747 & -0.747 & -0.747 & -0.746 & -0.818 & -0.818 & -0.817 & -0.752 \\ \hline
        germane & 3.542 & 3.542 & 3.518 & 3.542 & 3.542 & 3.542 & 3.543 & 3.518 & 3.518 & 3.518 & 3.542 \\ \hline
        magnesium chloride & -0.053 & -0.053 & -0.054 & -0.053 & -0.053 & -0.053 & -0.039 & -0.054 & -0.054 & -0.042 & -0.051 \\ \hline
        carbon oxyselenide & 1.368 & 1.368 & 1.364 & 1.368 & 1.368 & 1.368 & 1.369 & 1.364 & 1.364 & 1.366 & 1.368 \\ \hline
        titanium fluoride & -1.297 & -1.297 & -1.325 & -1.297 & -1.297 & -1.297 & -1.299 & -1.325 & -1.325 & -1.315 & -1.304 \\ \hline
        copper cyanide & -0.732 & -0.732 & -0.732 & -0.732 & -0.732 & -0.732 & -0.724 & -0.732 & -0.732 & -0.723 & -0.732 \\ \hline
        sodium tetramer & -0.461 & -0.461 & -0.462 & -0.461 & -0.461 & -0.461 & -0.452 & -0.462 & -0.462 & -0.448 & -0.471 \\ \hline
        hydrogen chloride & 2.928 & 2.928 & 2.925 & 2.927 & 2.928 & 2.928 & 2.930 & 2.925 & 2.925 & 2.926 & 2.933 \\ \hline
        arsine & 3.235 & 3.235 & 3.215 & 3.235 & 3.235 & 3.235 & 3.235 & 3.215 & 3.215 & 3.215 & 3.238 \\ \hline
        magnesium monoxide & -1.519 & -1.519 & -1.519 & -1.519 & -1.519 & -1.519 & -1.506 & -1.519 & -1.519 & -1.506 & -1.520 \\ \hline
        dipotassium & -0.295 & -0.295 & -0.296 & -0.295 & -0.295 & -0.295 & -0.290 & -0.296 & -0.295 & -0.288 & -0.310 \\ \hline
        sodium dimer & -0.206 & -0.206 & -0.206 & -0.206 & -0.206 & -0.206 & -0.198 & -0.206 & -0.206 & -0.199 & -0.230 \\ \hline
        chlorine & 0.006 & 0.006 & -0.006 & 0.006 & 0.006 & 0.006 & 0.007 & -0.006 & -0.006 & -0.006 & 0.006 \\ \hline
        adenine & 1.444 & 1.444 & 1.452 & 1.444 & 1.444 & 1.444 & 1.446 & 1.452 & 1.452 & 1.455 & 1.444 \\ \hline
        gallium monochloride & 0.333 & 0.333 & 0.309 & 0.333 & 0.333 & 0.333 & 0.336 & 0.309 & 0.309 & 0.311 & 0.332 \\ \hline
        hydrogen & 4.367 & 4.367 & 4.367 & 4.367 & 4.367 & 4.367 & 4.367 & 4.367 & 4.367 & 4.367 & 4.407 \\ \hline
        dirubidium & -0.336 & -0.336 & -0.336 & -0.336 & -0.336 & -0.336 & -0.324 & -0.336 & -0.336 & -0.324 & -0.343 \\ \hline
        silver dimer & -0.390 & -0.390 & -0.390 & -0.390 & -0.390 & -0.390 & -0.381 & -0.390 & -0.390 & -0.381 & -0.391 \\ \hline
        xenon & 7.738 & 7.738 & 7.728 & 7.738 & 7.738 & 7.738 & 7.740 & 7.728 & 7.728 & 7.729 & 7.738 \\ \hline
        iodine & -1.331 & -1.331 & -1.367 & -1.331 & -1.331 & -1.331 & -1.330 & -1.367 & -1.367 & -1.365 & -1.335 \\ \hline
        vinyl iodide & 1.676 & 1.676 & 1.665 & 1.676 & 1.676 & 1.676 & 1.678 & 1.665 & 1.665 & 1.666 & 1.674 \\ \hline
        aluminum iodide & -0.203 & -0.203 & -0.205 & -0.203 & -0.203 & -0.203 & -0.200 & -0.205 & -0.205 & -0.202 & -0.203 \\ \hline
        carbon tetraiodide & -1.403 & -1.403 & -1.427 & -1.403 & -1.403 & -1.403 & -1.402 & -1.427 & -1.427 & -1.427 & -1.407 \\
\bottomrule
\end{longtable}
\end{landscape}

\clearpage
\section{Wall-Clock Timings Water Clusters}
~
\begin{figure}[!h]
    \centering
    \includegraphics[width=0.9\linewidth]{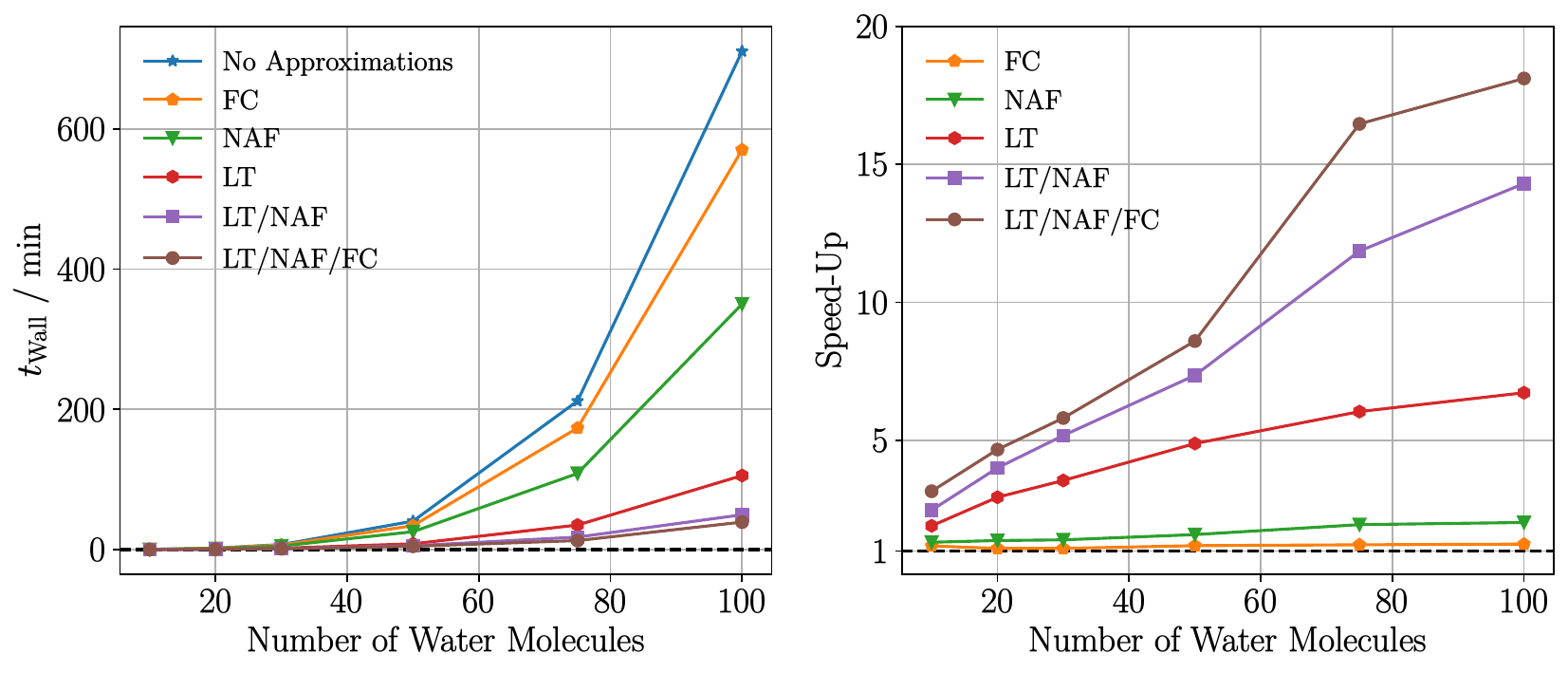}
    \caption{Computational wall-clock timings (left) and resulting speed-ups with respect to no approximations used (right) as a function of the number of water molecules of the AC-G$_0$W$_0$ calculations for the water clusters shown in the main text.}
    \label{fig:abs_timings}
\end{figure}
\begin{figure}[!h]
    \centering
    \includegraphics[width=0.87\linewidth]{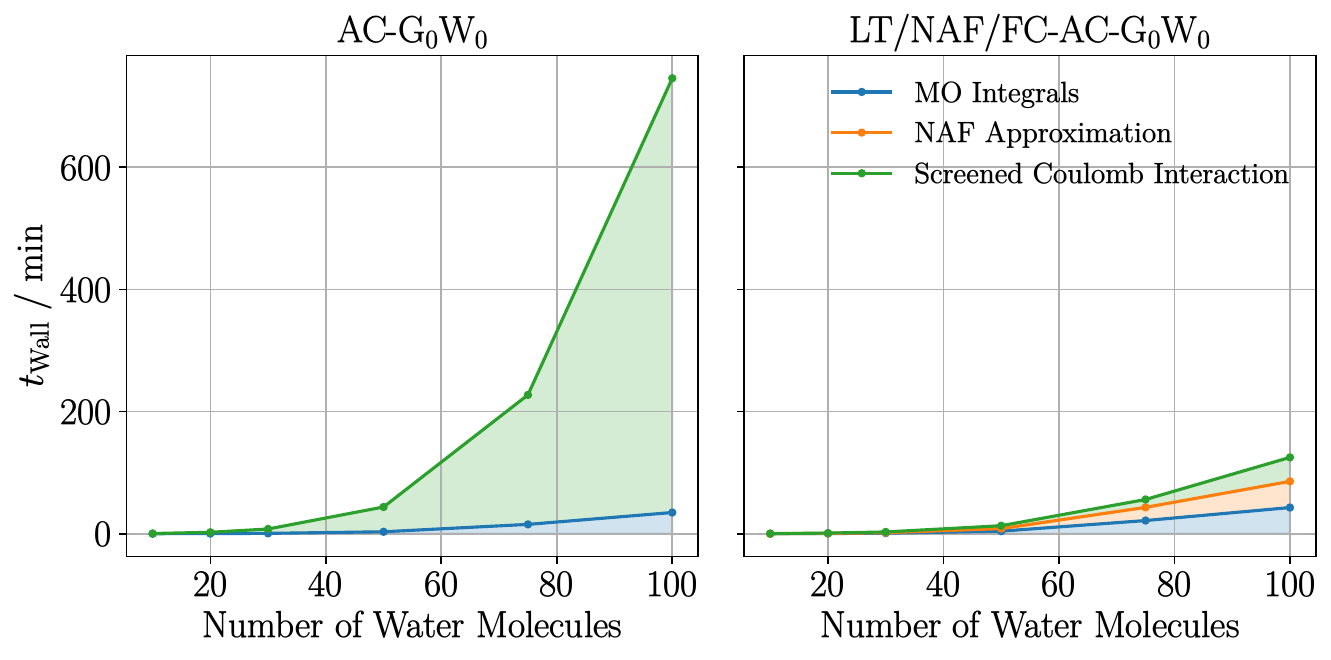}
    \caption{Composition of computational wall-clock timings of the regular AC-G$_0$W$_0$ (left) and LT/NAF/FC-AC-G$_0$W$_0$ calculations (right) for the water clusters shown in the main text. The graphs are understood as follows: The difference between the green line and the blue one (and not the green line itself) corresponds to the individual timing for the screened Coulomb interaction in the regular AC-GW case (implied with transparent green filling).}
    \label{fig:abs_timings}
\end{figure}

\begin{table}[H]
    \centering
    \caption{Computational wall-clock timings (in minutes) for different parts of G$_0$W$_0$ calculations for the water cluster containing 100 molecules. Listed are the preceding SCF calculation, the calculation of the screened Coulomb interaction $W$, the calculation of three-center molecular orbital (MO) integrals, and the natural auxiliary function (NAF) approximation. The calculations further applied a Laplace transformation and the frozen-core approximation. In parentheses are given the speed up with respect to the timing of the same category with half the cores used.}
    \begin{tabular}{lrlrlrlrlrl}
\toprule										
	$N_\mathrm{core}$	&	SCF	&	&$W$	&	&MO Integrals	&&	NAF	&\\
\midrule
	6	&	380	&	        &   196	&     &	271	&	&40	\\
	12	&	203 &    (1.9)	&	114	&(1.7)&	152	&(1.8)&	23	&(1.7)\\
	24	&	107	&    (1.9)  &	70	&(1.6)&	89	&(1.7)&	14	&(1.6)\\
	48	&	66	&    (1.6)  &	50	&(1.4)&	62	&(1.4)&	11	&(1.3)\\
	96	&	43	&    (1.5)  &	39	&(1.3)&	43	&(1.4)&	11	&(1.0)\\
\bottomrule
\end{tabular}
    \label{tab:thread_scaling}
\end{table}

\clearpage
\printbibliography